\documentclass[aps,prl,twocolumn,superscriptaddress]{revtex4-2}

\usepackage[latin1]{inputenc}
\usepackage{graphicx}
\usepackage{grffile}
\usepackage{amssymb}
\usepackage{amsmath}
\usepackage{bbold}
\usepackage{xspace}
\usepackage{dcolumn}
\usepackage{bm}
\usepackage{color}
\usepackage{float}
\usepackage[extra]{tipa}
\usepackage{MnSymbol}
\usepackage{braket}
\usepackage{dsfont} 
\usepackage{enumerate}
\usepackage[colorlinks=true,urlcolor=blue,linkcolor=blue,citecolor=blue,bookmarks=false]{hyperref}
\usepackage{soul}

\newcounter{myequation}
\makeatletter
\@addtoreset{equation}{myequation}
\makeatother

\newcounter{myfigure}
\makeatletter
\@addtoreset{figure}{myfigure}
\makeatother

\setcitestyle{square,numbers}

\DeclareMathOperator{\Tr}{Tr}

\newcommand{\ie}[0]{i.e.\@\xspace}
\newcommand{\eg}[0]{e.g.\@\xspace}

\newfont{\tensy}{cmsy10}

\newcommand{\ham}[1]{\hat{H}_{#1}}
\newcommand{\hamcl}[1]{{H}_{#1}}

\newcommand{\fan}[1]{\hat{c}^{\vphantom\dagger}_{#1}}
\newcommand{\fcr}[1]{\hat{c}^{\dagger}_{#1}}
\newcommand{\fantext}[1]{\hat{c}_{#1}}
\newcommand{\fden}[1]{\hat{n}_{#1}}

\newcommand{\Q}[1]{\hat{q}_{#1}}
\renewcommand{\P}[1]{\hat{p}_{#1}}

\newcommand{\q}[1]{q_{#1}}

\newcommand{\qvec}{\vec{q}}

\newcommand{\kB}{k_{\text{B}}}

\newcommand{\im}{\mathrm{i}}

\newcommand{\Hc}{\mathrm{H.c.}}
\newcommand{\absolute}[1]{\left| #1 \right|}

\newcommand{\expv}[1]{\left\langle #1 \right\rangle}
\newcommand{\expvtext}[1]{\langle #1 \rangle}

\newcommand{\expvc}[1]{\left\llangle #1 \right\rrangle_{\qvec}}

\newcommand{\expvctext}[1]{\llangle #1 \rrangle_{\qvec}}

\newcommand{\tave}{t_\mathrm{ave}}
\newcommand{\trel}{t_\mathrm{rel}}

\begin{document}


\title{Electronic Mechanism that Quenches Field-Driven Heating\\as Illustrated with the Static Holstein Model
}


\author{Manuel Weber}
\affiliation{\mbox{Department of Physics, Georgetown University, Washington,
    DC 20057, USA}}
\affiliation{\mbox{Max-Planck-Institut f\"ur Physik komplexer Systeme, N\"othnitzer Str.~38, 01187 Dresden, Germany}}
\affiliation{Institut f\"ur Theoretische Physik and W\"urzburg-Dresden Cluster of Excellence ct.qmat, Technische Universit\"at Dresden, 01062 Dresden, Germany}
\author{James K. Freericks}
\affiliation{\mbox{Department of Physics, Georgetown University, Washington,
    DC 20057, USA}}


\date{\today}




\begin{abstract}
Time-dependent driving of quantum systems has emerged as a powerful tool to engineer
exotic phases far from thermal equilibrium, but in the presence of many-body interactions it also leads to runaway heating, so that generic systems are believed to heat up until they reach a featureless infinite-temperature state. Understanding the mechanisms by which such a heat death can be slowed down or even avoided is a major goal---one such mechanism is to drive toward an even distribution of electrons in momentum space. Here we show how such a mechanism avoids runaway heating for an interacting charge-density-wave chain with a macroscopic number of conserved quantities when driven by a strong dc electric field; minibands with nontrivial distribution functions develop as the current is prematurely driven to zero.
Moreover, when approaching a zero-temperature resonance,
the field strength can tune between positive,
negative, or close-to-infinite effective temperatures for each miniband.
Our results suggest that nontrivial metastable distribution functions should
be realized in the prethermal regime of quantum systems coupled to slow bosonic modes.
\end{abstract}

\maketitle

The possibility to induce exotic nonequilibrium states with time-dependent electromagnetic fields in solid-state systems or in optical lattices has boosted the interest in driven quantum matter \cite{
2017NatMa..16.1077B}.
A current focus has been on Floquet systems
where a time-periodic drive can
realize novel topological phases
\cite{
2019ARCMP..10..387O,
2020NatRP...2..229R}
or time crystals \cite{PhysRevLett.116.250401,2018RPPh...81a6401S,2019arXiv191010745K}. Because time-dependent
Hamiltonians break energy conservation,
the presence of many-body interactions, like a coupling to a bath or to phonons,
inevitably leads to incoherent scattering and 
modifies the relaxation mechanisms of the electrons
 \cite{PhysRevX.4.041048, PhysRevE.90.012110}.
Under which circumstances the looming heat death can be delayed \cite{2015PhRvL.115y6803A, PhysRevX.7.011026, PhysRevB.93.155132, 2017CMaPh.354..809A, PhysRevX.10.021046} or even avoided \cite{PhysRevLett.115.030402, 2015PhRvL.114n0401P, 2016AnPhy.372....1A} in a driven many-particle system is an ongoing research topic that is of immediate importance for the experimental realization of novel out-of-equilibrium phases \cite{PhysRevX.10.021044, Peng:2021aa}. For instance, the breakdown of ergodicity in the many-body-localized phase \cite{2006AnPhy.321.1126B} has been considered as a microscopic process to avoid the heat death \cite{PhysRevLett.115.030402, 2015PhRvL.114n0401P, 2016AnPhy.372....1A}, but also in disorder-free realizations with a macroscopic number of conserved quantities \cite{PhysRevLett.118.266601}.
We study the nonequilibrium electron-phonon-coupled system, which remains too difficult to be solved exactly (for long times and large system sizes). Hence, one must make approximations that produce solutions in different limits. Here, we examine the case where the electrons interact with static phonons. This brings in limitations where heat is not directly transferred between the electrons and phonons. Nevertheless, any rapid processes occurring on electronic timescales should remain robust because once heating in the electronic system is quenched, adding energy exchange between electrons and phonons cannot significantly change the results.

In this Letter, we examine
periodically driven systems that do \textit{not} heat up indefinitely and study the logical follow-up questions:
What does the steady state look like and how is it reached as a function of time?
To this end, we consider a minimal interacting model where itinerant electrons on a chain are coupled to adiabatic phonons.
Starting from a thermal state, we drive our system with a dc electric field, representing the simplest realization of a Floquet system (due to Bloch oscillations).
This setup allows us to sample the initial states with a classical Monte Carlo method
and reach the steady state on lattice sizes much larger than in state-of-the-art exact-diagonalization studies. To characterize our final states,
we look at the frequency-resolved electron distribution function.
In thermal equilibrium, the occupation of states
is governed by the Fermi-Dirac distribution
$f_\mathrm{eq}(\omega) = 1/[\exp(\beta \omega) + 1]$
and only depends on the inverse temperature $\beta = 1/\kB T$.
The fluctuation-dissipation theorem relates $f_\mathrm{eq}(\omega)$
to the ratio of lesser and retarded single-particle Green's functions
(defined below).
In the same way, we define a nonequilibrium
distribution function $f_\mathrm{\infty}(\omega)$ for the steady state.
Only if our system reaches a thermal state will $f_\mathrm{\infty}(\omega)$  correspond to $f_\mathrm{eq}(\omega)$ with a renormalized temperature.
Our main results are shown in Fig.~\ref{fig:FT_spec_E1.0}.
The steady-state
spectral functions consist of minibands centered at integer multiples
of the electric field (due to the Wannier-Stark ladder formation).
For each miniband, we find Fermi-Dirac-like distribution functions with negative,
positive, or zero slope corresponding to positive, negative, or infinite
effective temperatures, respectively.
The cases with nontrivial distribution functions are highly nonequilibrium, because the distribution function should be a single one for all minibands, not a different one for each miniband;
the midpoints of
each miniband also follow a separate distribution function.
The proximity to the heat-death scenario can be tuned by adjusting the electric field close to a zero-temperature resonance that lifts Wannier-Stark localization.
Away from these points,
our system never fully heats up to infinite temperature;
importantly, we identify the symmetrization of the gauge-invariant momentum distribution function as the underlying mechanism to avoid the runaway heating.

\begin{figure}
  \includegraphics[width=0.9\linewidth]{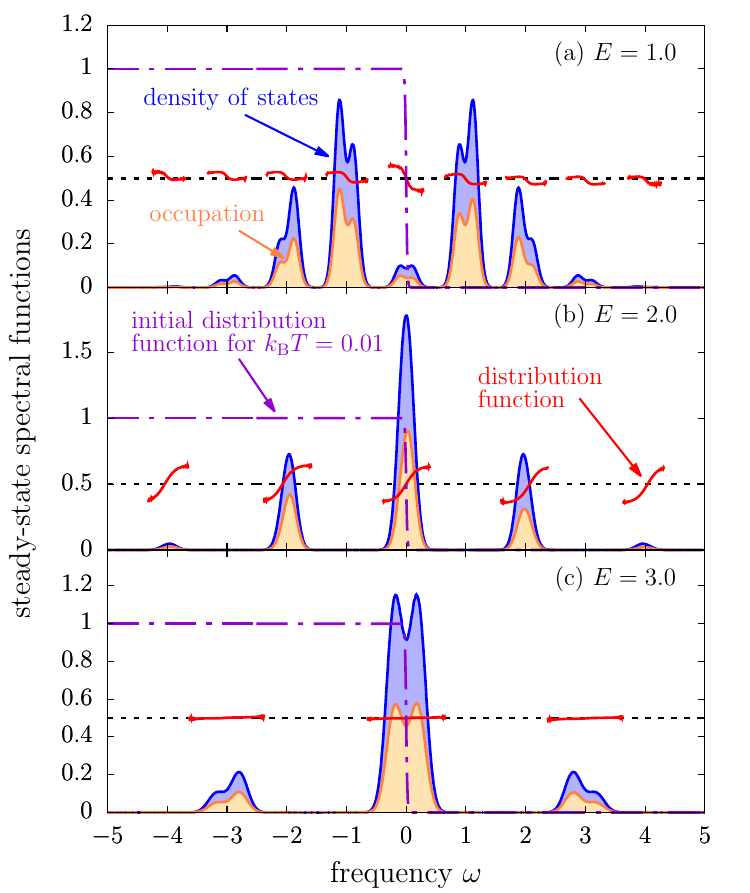}
  \caption{\label{fig:FT_spec_E1.0}%
Steady-state spectral functions.
Density of states, occupation, and distribution function for electric field strengths of (a) $E=1.0$, (b) $E=2.0$, and (c) $E=3.0$.
Here, $\kB T = 0.01$, $L=42$, $\lambda=0.5$.
  }
\end{figure}

To study the nontrivial properties of the steady state,
we consider the 1D Holstein model
$\hat{H}(t) = \hat{H}_\mathrm{el}(t) + \hat{H}_\mathrm{ph}$
in an electric field.
The electronic subsystem is given by
\begin{align}
\hat{H}_\mathrm{el}(t)
	=
	-J \sum_i  \left( e^{-\im \phi(t)} \fcr{i} \fan{i+1} + \Hc \right)
	+ g \sum_i \Q{i}  \left( \fden{i} - \mbox{$\frac{1}{2}$} \right).
\end{align}
The first term describes the nearest-neighbor hopping of spinless fermions
with amplitude $J$ where $\fcr{i}$ ($\fan{i}$) creates (annihilates) an electron at site $i$.
The time-dependent flux $\phi(t) = - E\, t \, \theta(t)$ incorporates a constant electric
field $E$
that is turned on at $t=0$.
We use the temporal gauge where $\hat{H}_\mathrm{el}(t)$ becomes a Floquet system with periodicity $2\pi/E$ induced by the periodic band structure.
In the second term, the local electron density $\fden{i} = \fcr{i} \fan{i}$ couples
to the phonon displacement $\Q{i}$.
The phonon Hamiltonian reads
$\hat{H}_\mathrm{ph} =  \sum_i ( \frac{K}{2} \Q{i}^2 + \frac{1}{2M} \P{i}^2)$
with stiffness constant $K$, mass $M$, and momentum $\P{i}$.
We define the dimensionless coupling $\lambda=g^2/4KJ$,
set $e=\hbar=c=1$, and fix $J=1$ as the unit of energy. All results are for $L=42$ sites with periodic boundary conditions.

In this Letter, we solve the real-time dynamics of $\hat{H}(t)$ exactly in
the adiabatic limit $M\to\infty$ of zero phonon frequency
where the phonons lose their dynamics and are unable to directly exchange energy with the electrons.
Then, the phonon displacements become classical variables
$\qvec = \{ \q{1}, \dots, \q{L} \}$
and their equilibrium distribution
\begin{align}
\label{weight_conf}
W_\mathrm{eq}[\qvec]
=
\frac{1}{Z}
e^{-\beta \hamcl{\mathrm{ph}}[\qvec]}
Z_{\mathrm{el}}[\qvec]
\end{align}
can be sampled using a Monte Carlo method
\cite{Michielsen1997, PhysRevB.94.155150}.
Any observable 
$
\expvtext{\hat{O}(t)}
=
\int d\qvec \, W_\mathrm{eq}[\qvec] \, \expvctext{\hat{O}(t)}
$
of the interacting system
reduces to a 
weighted average over noninteracting expectation values
\begin{align}
\label{obs_conf}
\expvc{\hat{O}(t)}
=
\frac{1}{Z_{\mathrm{el}}[\qvec]}
\Tr \left\{
e^{-\beta(\ham{\mathrm{el}}[\qvec]-\mu \hat{N})} \hat{O}_{\qvec}(t)
  \right\}
\end{align}
for a fixed  $\qvec$.
Here,
$Z_{\mathrm{el}} = \Tr \exp [-\beta(\ham{\mathrm{el}} - \mu \hat{N} )]$
is the partition function of the electronic subsystem with chemical potential $\mu$ and total electron number
$\hat{N}$.
While the phonons remain static, the electronic
subsystem
evolves according to the Heisenberg equations of motion
for $\fcr{i}(t) = \hat{U}^\dagger(t,t_0) \, \fcr{i}(t_0) \, \hat{U}(t,t_0)$.
Because 
$\hat{H}_{\mathrm{el}}(\qvec,t)$ is quadratic,
we only have to evolve the single-particle states using
a Trotter decomposition.
For a constant field $E$, the time-evolution operator
$\hat{U}(t,t_0)$ only needs to be calculated within its
period $\tau = 2\pi / E$.
For our simulations, we use the Trotter step $\Delta t = 2\pi / 3360\approx 0.002$ and calculate the steady-state behavior
at $1000 \, \tau$.
Note that, although the adiabatic limit excludes inelastic electron-phonon scattering because displacements $\qvec$ are conserved, the thermal phonon average recovers elastic electron-phonon scattering and therefore interaction effects.

We prepare our system in a thermal state
with initial temperature $\kB T$ and fix $\lambda=0.5$.
The phonon distribution $W_\mathrm{eq}[\qvec]$ is
entirely determined by $\kB T$.
At $\kB T=0$, the mean-field solution $\q{i} = (-1)^i \Delta / g$
is exact and leads to a band insulator with a single-particle
gap $\Delta\approx 0.3404$.
Translational symmetry is spontaneously 
broken by the periodic lattice distortion which gives rise to charge-density-wave order.
Many-body interactions are gradually incorporated with increasing $\kB T$, as electrons start to scatter elastically from thermally generated phonon displacements.
Already small fluctuations in the phonon fields
lead to a disordered phase, but the 
single-particle gap is only fully filled in at $\kB T \approx 0.1$,
where short-range charge-density-wave correlations disappear.
At higher temperatures, $W_\mathrm{eq}[\qvec]$ eventually becomes a Gaussian with a variance
$\sigma^2 \propto \kB T$.
For further details on the equilibrium solution, see Ref.~\cite{PhysRevB.94.155150}.

For a noninteracting system with a single band only, the application of a dc electric field
leads to Bloch oscillations with periodicity $2\pi/E$ in time-evolved observables like the electronic energy
$E_\mathrm{el}(t) = \expvtext{\hat{H}_\mathrm{el}(t)}/L$
or the current $j(t) = -J \sum_i \expvtext{\im \, e^{-\im \phi(t)} \fcr{i}(t) \fan{i+1}(t) + \Hc}/L$. For our clean two-band insulator at $\kB T=0$, interband Zener tunneling will also populate the initially unoccupied upper band. The combination of Zener tunneling and Bragg reflections leads to very irregular oscillations \cite{PhysRevLett.74.1831}.
For any finite initial temperature, the nonequilibrium dynamics is fundamentally different:
Then, a true steady state with constant energy and zero current is reached, as we see from the transient behavior of $E_\mathrm{el}(t)$ and $j(t)$ in Figs.~\ref{fig:FT_equal_time}(a) and \ref{fig:FT_equal_time}(b), respectively.
\begin{figure}
  \includegraphics[width=0.9\linewidth]{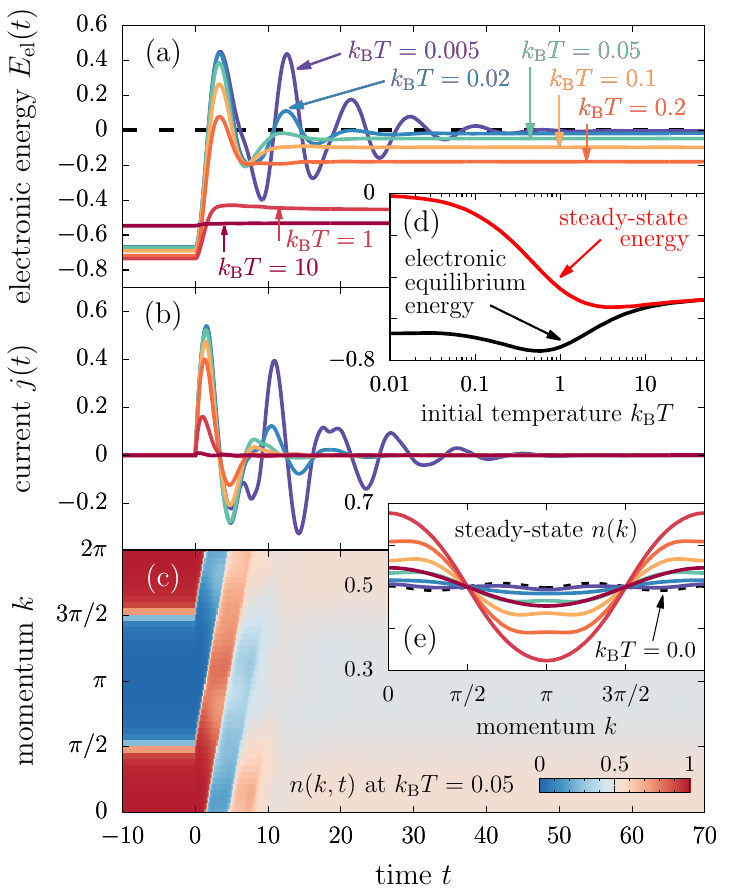}
  \caption{\label{fig:FT_equal_time}%
Transient nonequilibrium dynamics:
(a) Electronic energy and (b) current as a function of time for different initial temperatures.
The dashed line in (a) represents the time average of $E_\mathrm{el}(t)$ at $\kB T=0$.
(c) Gauge-invariant momentum distribution function at $\kB T = 0.05$.
(d) Comparison of the equilibrium and steady-state electronic energies
as a function of temperature.
(e) Steady-state momentum distribution function for different temperatures.
The labels in (a) also apply to (b) and (e).
Here, $E=1.0$, $L=42$, $\lambda=0.5$.
  }
\end{figure}
The damping of the average energy and current
results from the destructive interference between oscillating solutions for different phonon configurations.
The gauge-invariant momentum distribution function \cite{PhysRevB.44.3655}
$n(k,t) = \expvtext{ \fcr{{k+\phi(t)}}(t) \, \fan{{k+\phi(t)}}(t) }$
in Fig.~\ref{fig:FT_equal_time}(c) shows how the current
vanishes before the system can reach an infinite-temperature state. 
The momentum distribution becomes a nontrivial even function of $k$ in the long-time limit.
This points toward a restoration of time-reversal symmetry
in the steady state
as the current is simultaneously quenched.
A comparison of
$E_\mathrm{el}$
between initial and final states
in Fig.~\ref{fig:FT_equal_time}(d) reveals that
heating effects are strongest at low $\kB T$ where
the steady state gets close to the infinite-temperature
result $E_\mathrm{el}=0$. 
Surprisingly, a higher initial temperature
reduces the final energy and thereby the effective temperature of the steady state; this is similar to the inverse Mpemba effect~\cite{mpemba}.
In addition, the steady-state occupation $n(k)$ in Fig.~\ref{fig:FT_equal_time}(e)
is close to a uniform distribution at low $\kB T$ and reaches its strongest $k$ dependence around $\kB T =1$. It appears that the proximity to coherent bands at low $\kB T$ allows for stronger heating, whereas localization effects due to phonon-induced disorder steadily reduce the system's ability to absorb energy with increasing $\kB T$.
Note that $E_\mathrm{el}$
does not reach zero for $\kB T \to \infty$, neither in equilibrium nor for the steady state,
because the variance of the phonon distribution scales as $\kB T$ for large temperatures.

The spectral properties of the steady state can be inferred from the retarded and
lesser Green's functions
\begin{gather}
G^\mathrm{ret}_{ij}(t,t')
	=
	-\im \Theta(t - t') \expv{\left\{ \fan{i}(t), \fcr{j}(t') \right\}}
	\, , \\
G^{<}_{ij}(t,t')
	=
	\im  \expv{\fcr{j}(t') \, \fan{i}(t) }
	\, .
\end{gather}
Using the Wigner coordinates $\tave = (t + t')/2$ and $\trel = t - t'$,
we define the
Fourier transform
$
G^{\alpha}_\mathrm{loc}(\tave, \omega)
	=
	\int d\trel \,
	e^{\im(\omega+\im \eta)\trel} 
	\sum_i
	G^{\alpha}_{ii}(\tave+\trel/2,\tave-\trel/2) / L
$
of the local Green's functions.
Then, the density of states becomes
$A(\tave,\omega) = - \mathrm{Im} \, G^\mathrm{ret}_\mathrm{loc}(\tave,\omega) / \pi$
and the occupation
$A^{<}(\tave,\omega) = \mathrm{Im} \, G^<_\mathrm{loc}(\tave,\omega) / 2\pi$.
The steady-state spectra are shown in Fig.~\ref{fig:FT_spec_E1.0} for $\kB T =0.01$.
Their ratio defines the nonequilibrium distribution function,
\begin{align}
f_\infty(\omega) = \frac{A^{<}(\tave \to \infty,\omega)}{A(\tave \to \infty,\omega)} \, ,
\end{align}
which can be interpreted as a generalized nonequilibrium fluctuation-dissipation theorem in the long-time limit.

\begin{figure}
  \includegraphics[width=0.9\linewidth]{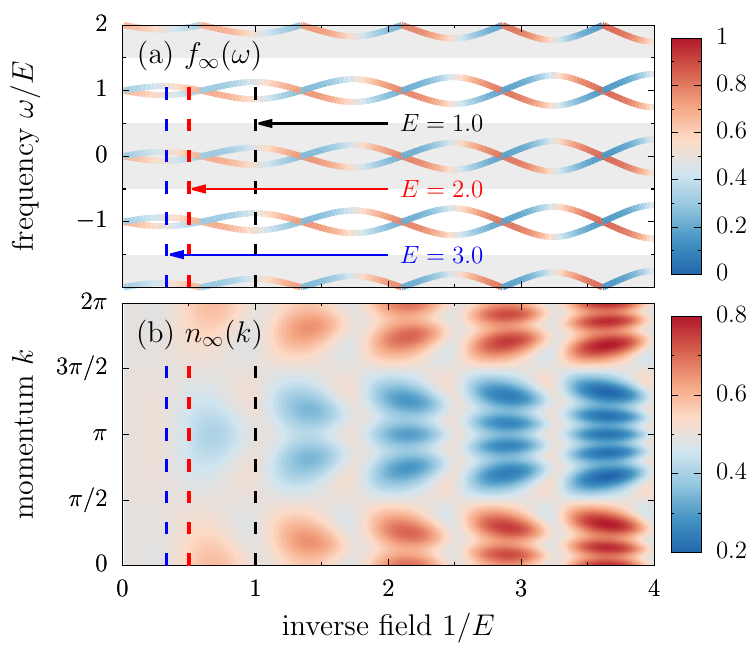}
\caption{\label{fig:phase_diagram}%
Steady-state distribution functions for zero initial temperature.
(a)~The two quasienergies per Floquet energy window show (anti)crossings as a function of inverse field. The color
coding corresponds to the spectral distribution function $f_\infty(\omega)$.
(b)~The momentum distribution function $n_\infty(k)$ becomes flat when $f_\infty(\epsilon_\nu)=1/2$.
Dashed lines indicate the parameters chosen in Figs.~\ref{fig:FT_spec_E1.0} and \ref{fig:FT_dist}.
Here $\lambda=0.5$.
  }
\end{figure}

We can understand the main spectral features  in Fig.~\ref{fig:FT_spec_E1.0}
from the zero-temperature limit.
Because of the doubling of the unit cell by the Peierls distortion,
the energy spectrum of the steady state in Fig.~\ref{fig:phase_diagram}(a) consists of two interpenetrating
Wannier-Stark ladders with a level spacing of $E$ each.
The color coding of the energy levels corresponds to $f_\infty(\omega)$,
which we calculate using Floquet theory. Because the zero-temperature Green's functions
do not decay with time, we average the spectra over $\tave$; in this way, steady-state observables are defined consistently at $\kB T=0$ and
$\kB T > 0$.
We obtain $f_\infty(\epsilon_\nu + m E) = \frac{2}{L} \sum_{p} \expvtext{\fden{p\nu}(t=0)}$
for $\epsilon_{1,2} \in [-E/2,E/2]$, independent of $m\in \mathds{Z}$.
Here, $\fden{p\nu}(t=0)$ is the number operator in the Floquet basis
with momentum $p\in [0,\pi)$. Hence, $f_\infty(\omega)$ is given by the
overlap of the Floquet states with the initially occupied states.
Within each Floquet energy window in Fig.~\ref{fig:phase_diagram}(a),
we find intervals of $E$ where the lower (upper) band has a higher $f_\infty$
corresponding to an effective positive (negative) temperature per miniband
in Fig.~\ref{fig:FT_spec_E1.0}(a)  [Fig.~\ref{fig:FT_spec_E1.0}(b)].
The two regimes are separated by a level crossing in the zone center as well
as an avoided level crossing at the zone boundary. Zener tunneling at
the avoided crossings lifts the Wannier-Stark localization
and leads to an
equal occupation of the two levels corresponding to an 
effective infinite
temperature in Fig.~\ref{fig:FT_spec_E1.0}(c).
At these resonances, the time-averaged gauge-invariant momentum
distribution function $n_\infty(k)$
is exactly $1/2$ for all $k$, as shown in Fig.~\ref{fig:phase_diagram}(b) and proved in the Supplemental Material
\footnote{See Supplemental Material at \url{url} for details on the zero-temperature solution, additional results, Refs.~\cite{McCoyWu+1973,RevModPhys.75.1333,PhysRevLett.122.247402,2020arXiv200308252R,PhysRev.138.B979,PhysRevLett.122.130604}, and data files for the results presented in this Letter.}.
When $1/E$ is tuned off resonance, $n_\infty(k)$
increasingly gains structure with each resonance that is crossed.
Resonance-induced delocalization is a well-known feature of coupled Wannier-Stark ladders \cite{Leo_1989,PhysRevLett.74.1831}
and has been observed experimentally, \eg,  in
semiconductor superlattices \cite{PhysRevLett.65.2720}.

\begin{figure}
  \includegraphics[width=0.9\linewidth]{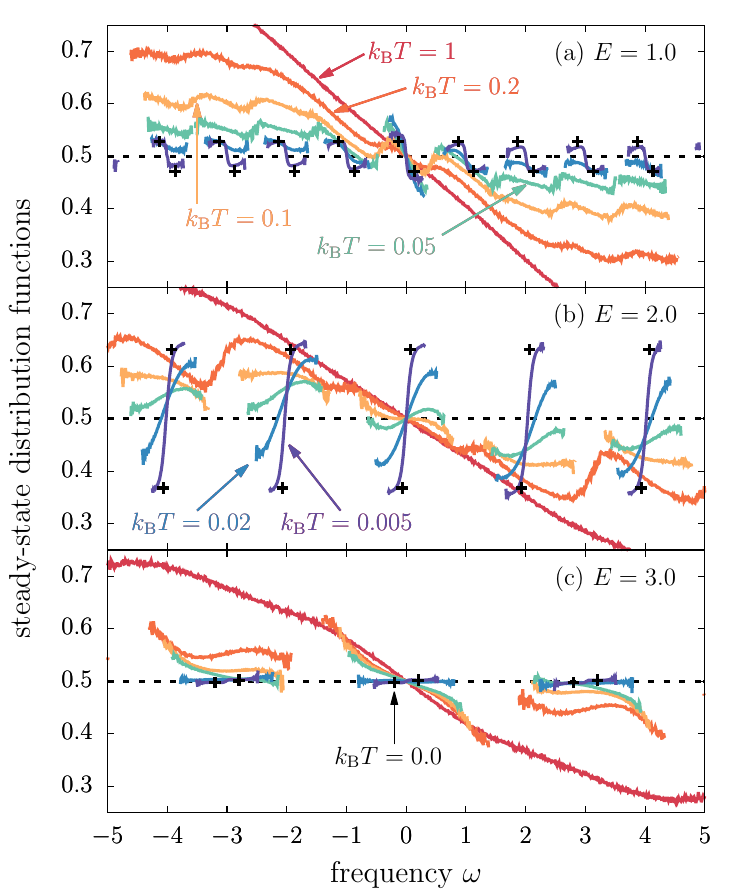}
  \caption{\label{fig:FT_dist}%
Steady-state distribution functions for different initial temperatures
and electric field strengths of (a) $E=1.0$, (b) $E=2.0$, and (c) $E=3.0$.
Here, $L=42$ and $\lambda=0.5$.
  }
\end{figure}

By introducing thermal fluctuations into the Floquet system via a nonzero initial $k_BT$, we can explain the spectral properties of Fig.~\ref{fig:FT_spec_E1.0}.
The phonon disorder lifts the $L/2$-fold degeneracy of each Floquet level
such that the delta peaks in the spectra get broadened.
Then, we can obtain
$f_\infty(\omega)$
on a continuous interval
around the original levels
as long as the spectral weight is not too small.
We study the effect of the initial temperature on $f_\infty(\omega)$
in greater detail in Fig.~\ref{fig:FT_dist}.
While the distributions per miniband
mainly get smeared out for
positive effective temperatures in Fig.~\ref{fig:FT_dist}(a),
increasing phonon fluctuations reverse the negative-temperature
distributions as a function of $\kB T$ in Fig.~\ref{fig:FT_dist}(b).
The flat distributions in Fig.~\ref{fig:FT_dist}(c) remain rather flat
for a broad range of $\kB T$.
Moreover,
the phonon fluctuations lift the degeneracy of
$f_\infty(\omega)$ between the different Floquet zones
such that the focal points of each miniband follow an overall
distribution function.
As $\kB T$ increases,
the latter slowly
transforms into a Fermi-Dirac-like distribution with an
effective temperature that decreases. Above $\kB T \approx 1.0$,
$f_\infty(\omega)$ is close to the initial thermal distribution and its effective
temperature increases again,
as suggested by the
steady-state energy in Fig.~\ref{fig:FT_equal_time}(d).

We can interpret the interplay between initial temperature and electric field
in terms of competing localization mechanisms.
At high temperatures, the strong Gaussian phonon disorder promotes Anderson localization.
Then, the application of an electric field
enhances the localization length \cite{Prigodin1980},
which only leads to small heating when approaching the steady state.
Hence, $f_\infty(\omega)$ becomes flatter with increasing $E$, as we see in Fig.~\ref{fig:FT_dist} for $\kB T = 1$.
Only if $E$ is strong enough compared to the phonon disorder do signatures of Wannier-Stark localization appear, as its
localization length is proportional to $1/E$ \cite{PhysRevB.35.8929}.
Therefore, the steady-state features at $\kB T \ll 1$ are governed by the Floquet solution. While each Floquet zone
is populated equally at $\kB T=0$, the nontrivial overall distribution for $\kB T > 0$ seems to be a partial memory effect of $f_\mathrm{eq}(\omega)$.
We saw that heating effects are strongest at low $\kB T$, where the system becomes a coherent band insulator.
Although observables at zero temperature never decay toward a true steady state,
a time average over all $t>0$ is consistent with the steady-state results at $\kB T \to 0$, as shown in Fig.~\ref{fig:FT_dist} for $f_\infty(\omega)$, in Fig.~\ref{fig:FT_equal_time}(a) for $E_\mathrm{el}$, 
or in Fig.~\ref{fig:FT_equal_time}(e) for $n_\infty(k)$ (for further data, see the Supplemental Material).
The higher absorption of heat at low $\kB T$ is thus determined by an easier ability for the system to equally occupy all electronic states as time proceeds.

In conclusion, we demonstrated for a simple
interacting
model of itinerant electrons coupled to adiabatic phonons that the application of a dc electric field
does
not lead to a featureless infinite-temperature state, unless the system is tuned to a zero-temperature resonance.
Instead, the heating of the electronic subsystem stops as the current is prematurely driven to zero due to the symmetrization of the momentum distribution function.
We obtain strongly nonequilibrium steady states with Fermi-Dirac-like distribution functions for each Floquet miniband. These
distribution functions can be tuned by the strength of the electric field, from positive to negative effective temperatures. 

It remains open how quantum lattice fluctuations further affect these findings. The adiabatic phonon limit is special in the sense that electrons can only scatter elastically off the static phonon displacements.
Inelastic scattering becomes important at timescales proportional to the inverse phonon frequency; for earlier times, the dynamics will be determined by the adiabatic phonon limit. For typical charge-density-wave systems, the phonons are (by several orders of magnitude) slower than the electrons. Therefore, the electron dynamics in Fig.~\ref{fig:FT_equal_time} has already reached a steady state for all but the lowest temperatures before realistic phonon dynamics can set in;
once the current is driven to zero via the symmetrized  momentum distribution, it is unclear what could destroy this for later times. Therefore, we expect the nonequilibrium distribution functions found in this Letter to still occur for low phonon frequencies and high $\kB T$ \footnote{In equilibrium, the adiabatic-phonon description is valid for temperatures much larger than the phonon frequency, as confirmed by exact quantum Monte Carlo simulations \cite{PhysRevB.98.235117}.}, at least in a long-lived transient regime, where elastic scattering is the dominant mechanism.
Although we have identified the symmetrization of the gauge-invariant momentum distribution function in a special setup, it will be worth studying how this mechanism affects heating in more complicated driven electron systems.

\begin{acknowledgments}
\textit{Acknowledgments.} 
 We acknowledge helpful discussions with A.~Kemper and D.~Luitz.
 This work was supported by the U.S. Department of Energy (DOE),
  Office of Science, Basic Energy Sciences (BES) under Award  DE-FG02-08ER46542. 
  Work at TU Dresden was supported by the Deutsche Forschungsgemeinschaft through the Würzburg-Dresden Cluster of Excellence on Complexity and Topology in Quantum Matter -- \textit{ct.qmat} (EXC 2147, Project No. 390858490).
  J.K.F.~was also supported by the McDevitt bequest at Georgetown University.
  The authors gratefully acknowledge the Gauss Centre for Supercomputing e.V. (www.gauss-centre.eu) for funding this project by providing computing time on the GCS Supercomputer SuperMUC-NG at Leibniz Supercomputing Centre (www.lrz.de) (Project-ID pr53ju).
\end{acknowledgments}


%

\clearpage

\stepcounter{myequation}
\stepcounter{myfigure}

\setcounter{secnumdepth}{3}  

\renewcommand{\thefigure}{S\arabic{figure}}
\renewcommand{\thesection}{S\arabic{section}}
\renewcommand{\thetable}{S\arabic{table}}
\renewcommand{\theequation}{S\arabic{equation}}

\definecolor{Gray}{gray}{0.9}

\onecolumngrid

\centerline{\bf\large Supplemental Material} \vskip5mm
\centerline{\bf\large for} \vskip5mm
\centerline{\bf\large Electronic Mechanism that Quenches Field-Driven Heating} \vskip0.75mm
\centerline{\bf\large as Illustrated with the Static Holstein Model} \vskip1cm

\twocolumngrid

\author{Manuel Weber}
\affiliation{Max-Planck-Institut f\"ur Physik komplexer Systeme, N\"othnitzer Str.~38, 01187 Dresden, Germany}
\affiliation{Institut f\"ur Theoretische Physik and W\"urzburg-Dresden Cluster of Excellence ct.qmat, Technische Universit\"at Dresden,
01062 Dresden, Germany}
\author{Matthias Vojta}
\affiliation{Institut f\"ur Theoretische Physik and W\"urzburg-Dresden Cluster of Excellence ct.qmat, Technische Universit\"at Dresden,
01062 Dresden, Germany}

\date{\today}

\maketitle

\section{Many-body physics, thermalization and static models}

Static models like the Ising model, the Falicov-Kimball model, or the static Holstein model are solved by using an annealed average over the statistical ensemble of possible values for the static variables. While it may seem like this averaging over instantiations of particular distributions of the static coordinate will remain with the same behavior as the noninteracting nature of each instantiation, this is not true. 
Quantum fluctuations are suppressed, but not statistical fluctuations. The statistical fluctuations are enough to make the statistical ensemble behave as a nontrivial \textit{interacting} system that includes many of the effects seen in non-static many-body models.  

Let us first show this by considering the Ising model in two dimensions \cite{McCoyWu+1973}.
The system is in its eigenstate basis when we express the states as product states in the direction of the Ising interaction. This is the equivalent of a set of noninteracting quantum states, which, for each element of the average, satisfy all the properties of noninteracting systems. In particular, the model does not allow any spin flips. But, after performing a statistical average and using the laws of statistical mechanics, we find this average over noninteracting systems has a finite-temperature phase transition with nontrivial (in the sense of non mean-field theory and hence generic) critical exponents. How can this be? Because the statistical averaging over the different noninteracting components provides a system that is interacting and has nontrivial behavior that is not just that of each noninteracting component of the average. 

In electronic systems, the analog of this is the Falicov-Kimball model \cite{RevModPhys.75.1333}. Again, all of the similar statements can be made. Each configuration is noninteracting, so the average must be too. But just like for the Ising model, this is incorrect. The Falicov-Kimball model has an exact solution in infinite dimensions and that solution has (i) the Mott transition; (ii) order-disorder transitions with finite transition temperatures; (iii) a dynamical self energy, that reacts on fast time scales, even though it interacts with static scatterers; and (iv) all of the generic dynamical effects of strongly correlated materials, including a proper generic behavior of the optical conductivity. What is it missing? It does not describe a Fermi-liquid phase at low temperatures. Those are driven by quantum effects absent in the model when interactions are turned on. But, because the Fermi-liquid temperature is renormalized to zero as the Mott transition is approached, this regime is limited to low temperature and weak interacting regions only. 

Now, we come to the electron-phonon coupled model, which we consider in the adiabatic limit of infinite ion mass. Based on these two examples we expect this model to illustrate the following behavior---(i) it is an interacting model due to the statistical averaging, not a noninteracting model; (ii) it has a dynamical self-energy that includes fast dynamics of the electrons, even if the scattering centers are static; (iii) it has order-disorder transitions to charge-density wave physics (these are suppressed to only the ground state in one dimension); and (iv) the behavior is expected to be generic outside the quantum coherent regime (meaning it will be accurate at high temperatures, at steady states with high average energy, and when the correlations are strong). What does it not have? It has no Fermi-liquid state at low temperature or Luttinger-liquid state in one dimension. It also does not support superconducting solutions. But, we do expect these types of models will display generic many-body physics behavior in the regimes we are examining. 

And what about thermalization? The static quantities are conserved by the Hamiltonian, so they do not relax. These constraints often lead to generalized Gibbs distributions rather than equilibrium distributions, when the system relaxes. But, it is often the case when a system is pumped, that the generalized Gibbs ensemble remains close to the equilibrium distributions \cite{PhysRevLett.122.247402} and in the case considered here, where we drive to a nonequilibrium steady state, we expect the properties of the steady state to be similar to what would be found in models that are not static, but precisely how close is not yet known.

\section{Floquet theory for the perfectly-dimerized chain}

At zero initial temperature, the Holstein model driven by a constant electric field can be solved efficiently using Floquet theory. Here, we want to give a brief introduction to Floquet theory 
and outline the relevant steps for our solution, before we present additional results in the
subsequent section. Our presentation of the basics of Floquet theory follows Ref.~\cite{2020arXiv200308252R}.

\subsection{Basics of Floquet theory}

For a time-dependent Hamiltonian $\hat{H}(t+T) = \hat{H}(t)$ with
periodicity $T$, the Schr\"odinger equation ($\hbar = 1$)
\begin{align}
\label{Eq:Schrodinger}
\im \frac{d}{dt} \ket{\Psi(t)} = \hat{H}(t) \ket{\Psi(t)}
\end{align}
can be solved using Floquet theory.
The eigenstates of the time-evolution operator
after evolving by one time period, $\hat{U}(t_0+T,t_0)$, are the so-called \textit{Floquet states}.
According to Floquet's theorem, these states can be expanded in terms of time-periodic states $|\Phi_\nu(t)\rangle$ via
\begin{align}
\label{Eq:State_Floquet}
\ket{\Psi_\nu(t)}
	=
	e^{-\im \epsilon_\nu t} \ket{\Phi_\nu(t)} \, , \quad
\ket{\Phi_\nu(t+T)} = \ket{\Phi_\nu(t)} \, ,
\end{align}
with $\epsilon_\nu$ the Floquet quasienergies.
Because $\ket{\Phi_\nu(t)}$ has periodicity $T$, we can
expand these states in a Fourier series with frequency
$\Omega = 2\pi / T$,
\begin{align}
\label{Eq:State_Floquet_FS}
\ket{\Phi_\nu(t)} = \sum_{m=-\infty}^\infty e^{-\im m \Omega t} \ket{\phi_\nu^{(m)}} \, .
\end{align}
Here, $\ket{\phi_\nu^{(m)}}$ is the $m$-th Fourier coefficient of $\ket{\Phi_\nu(t)}$.
If we plug Eq.~(\ref{Eq:State_Floquet}) and Eq.~(\ref{Eq:State_Floquet_FS})
into Eq.~(\ref{Eq:Schrodinger})
and expand the Hamiltonian as
\begin{align}
\label{Eq:HamFloquetSeries}
\hat{H}(t) = \sum_{m=-\infty}^\infty e^{-\im m \Omega t} \hat{H}^{(m)} \, ,
\end{align}
we obtain the eigenvalue equation
\begin{align}
\label{Eq:eigenval}
 \left( \epsilon_\nu + m \Omega \right) \ket{\phi_\nu^{(m)}} 
=
\sum_n \hat{H}^{(m-n)} \ket{\phi_\nu^{(n)}} \, .
\end{align}
To determine $\epsilon_\nu$ and $\ket{\phi_\nu^{(m)}}$, 
we only need to diagonalize the infinite-dimensional matrix
\begin{align}
\label{Eq:FloquetHam}
\hat{H}^\mathrm{F}_{mn} =\hat{H}^{(m-n)} - m \Omega \, \delta_{mn} \, .
\end{align}
For the numerical diagonalization of $\hat{H}^\mathrm{F}$, we introduce a cutoff that is large enough so that results are converged.
A complete set of states is found by restricting
all quasienergies to lie in the range $-\Omega/2 \leq \epsilon_\nu < \Omega/2$.
For further details see Ref.~\cite{2020arXiv200308252R} or the original work \cite{PhysRev.138.B979}.

\subsection{Holstein model at zero temperature}

At zero temperature and half-filling, the phonon displacements in the Holstein model
are perfectly dimerized and described by the mean-field ansatz $\q{i} = (-1)^i \Delta / g$.
Then, the Holstein model in a constant electric field $E$ can be partially diagonalized and we obtain
the time-dependent two-band Hamiltonian
\begin{align}
\hat{H}(t)
	=
	\sum_{p} \sum_{\alpha \beta}
	\fcr{p\alpha} \, {\mathcal{H}}_{p,\alpha \beta}(t) \, \fan{p\beta} \, .
\end{align}
Here, $\hat{c}_{p\alpha} = \hat{c}_{p+(\alpha-1) \pi}$ annihilates an electron
with reduced momentum $p=2\pi n/L$, $n=\{0, \dots, L/2 -1\}$,
and orbital index $\alpha=\{1,2\}$.
The single-particle
Hamiltonian is 
\begin{align}
\hat{\mathcal{H}}_p(t)
	=
\begin{pmatrix}
\epsilon(p+Et) & \Delta  \\
\Delta & -\epsilon(p+Et) 
\end{pmatrix} \, .
\end{align}
where $\epsilon(p) = -2J \cos p$.
Because
$\hat{\mathcal{H}}_p(t)$ has periodicity $T=2\pi / E$,
we can expand it according to Eq.~(\ref{Eq:HamFloquetSeries})
with $\Omega = 2\pi / T = E$. The only nonzero elements are
\begin{align}
\label{Eq:Floquet_Matrix_HS}
\hat{\mathcal{H}}_p^{(0)}
	=
\begin{pmatrix}
0 & \Delta  \\
\Delta & 0
\end{pmatrix}
\, ,
\qquad
\hat{\mathcal{H}}_p^{(\pm 1)}
	=
\begin{pmatrix}
-J e^{\mp \im p} & 0  \\
0 & J e^{\mp \im p}
\end{pmatrix}
\, .
\end{align}
With this, the eigenvalue equation (\ref{Eq:eigenval}) can be solved for each
momentum $p$.
Note that we have an additional orbital index. The Fourier expansion
of the Floquet states can be represented as
\begin{align}
\ket{\phi_{p\nu}^{(m)}} =
\sum_\alpha \braket{p\alpha | \phi_{p\nu}^{(m)}}  \ket{p\alpha} \, ,
\end{align}
where  $\ket{p\alpha}$ represents the eigenbasis of the physical states
created by the corresponding creation operators.

\subsubsection{Symmetries of the Floquet Hamiltonian}

For the 1D Holstein model in a
constant field,
we  want to give a few symmetries of
the infinite-dimensional
(Floquet) Hamiltonian that simplify its solution:
\begin{enumerate}[(i)]
\item The \textit{momentum translation symmetry} 
\begin{align}
\hat{S}_p^\dagger \, \hat{\mathcal{H}}^\mathrm{F}_p \,  \hat{S}_p
 = 
\hat{\mathcal{H}}^\mathrm{F}_{p=0} \, ,
\quad 
\hat{S}_{p,mn} = e^{-\im p m} \delta_{mn}
\\
\Rightarrow \quad
\braket{p\alpha | \phi_{p\nu}^{(m)}}
 = 
e^{-\im p m}
 \braket{p=0,\alpha | \phi_{p=0,\nu}^{(m)}} 
\label{Eq:State_p_transf}
\end{align}
relates all expectation values to the $p=0$ case. In particular, the Floquet quasienergies
become independent of $p$, \ie, $\epsilon_{p\nu} = \epsilon_\nu$.

\item The \textit{orbital (sublattice interchange) symmetry}
\begin{align}
\hat{S}_p^\dagger \, \hat{\mathcal{H}}^\mathrm{F}_p \,  \hat{S}_p
 = 
\hat{\mathcal{H}}^\mathrm{F}_{p} \, ,
\quad 
\hat{S}_{p,mn} = e^{-\im \pi m} \hat{\sigma}_x \, \delta_{mn}
\\
\Rightarrow \quad
\braket{p\alpha | \phi_{p\nu}^{(m)}}
= 
e^{-\im \pi m}
\braket{p \bar{\alpha} | \phi_{p\nu}^{(m)}}
\label{Eq:State_subl_transf}
\end{align}
relates opposite orbitals $\alpha$ and $\bar{\alpha}$. Here, $\hat{\sigma}_x$ is the usual Pauli matrix.

\item The \textit{particle-hole symmetry}
\begin{align}
\hat{S}_p^\dagger \, \hat{\mathcal{H}}_p \,  \hat{S}_p
 = 
- \hat{\mathcal{H}}_{p} \, ,
\quad 
\hat{S}_{p,mn} = e^{-\im \pi m} \hat{\sigma}_z \, \delta_{mn}
\\
\Rightarrow \quad
\braket{p\alpha | \phi_{p\nu}^{(m)}}
= 
e^{-\im \pi m}
\braket{p {\alpha} | \phi_{p\bar{\nu}}^{(-m)}}
\label{Eq:State_ph_transf}
\end{align}
relates $\epsilon_\nu + m\Omega \to - \epsilon_\nu - m \Omega$.
\end{enumerate}

\subsubsection{Equilibrium solution}

We prepare the initial state of our system in the half-filled ground state of the
equilibrium Hamiltonian with $E=0$. Below, we need
$
\expvtext{\fcr{p\alpha }\fantext{p\beta}}
$,
which can be obtained from diagonalizing the Hamiltonian for each $p$
and filling the lower level. We obtain
\begin{align}
\expv{\fcr{p2} \fan{p 2}}
=
\frac{\Delta^2}
{
2 \left[\Delta^2 + \epsilon(p)^2 - \epsilon(p) \sqrt{\Delta^2 + \epsilon(p)^2}\right]
} \, ,
\\
\expv{\fcr{p1} \fan{p 2}}
=
\frac{\Delta \left[ \epsilon(p) - \sqrt{\Delta^2 + \epsilon(p)^2}  \right]}
{
2 \left[\Delta^2 + \epsilon(p)^2 - \epsilon(p) \sqrt{\Delta^2 + \epsilon(p)^2}\right] 
} \, ,
\end{align}
as well as the relations
\begin{align}
\expv{\fcr{p1} \fan{p 1}}
=
1 - \expv{\fcr{p2} \fan{p 2}} \, ,
\qquad
\expv{\fcr{p2} \fan{p 1}}
= \expv{\fcr{p1} \fan{p 2}} \, .
\end{align}

\subsection{Time-evolution operator in the Floquet basis}

To calculate real-time observables, we need access
to the time-evolved creation and annihilation operators. Using the equation
of motion, we can trace the time evolution of any quadratic Hamiltonian back to an initial time $t_0$, \ie,
\begin{align}
\nonumber
\fan{p\alpha}(t)
	&=
	\hat{U}^\dagger(t,t_0) \, \fan{p\alpha}(t_0) \, \hat{U}(t,t_0) \\
	&= \sum_{\alpha'} {\mathcal{U}}_{p,\alpha\alpha'}(t,t_0) \,  \fan{p\alpha'}(t_0) \, .
\label{Eq:EOM}
\end{align}
The  time-evolution operator of the single-particle Hamiltonian can be obtained as
\begin{align}
 {\mathcal{U}}_{p,\alpha\alpha'}&(t,t_0) 
	=
	\bra{{p\alpha}} \hat{\mathcal{U}}(t,t_0) \ket{{p\alpha'}}
	\nonumber \\
	&=
	\sum_{\nu} \bra{{p\alpha}} \hat{\mathcal{U}}(t,t_0) \ket{ \Psi_{p\nu}(t_0)}\braket{ \Psi_{p\nu}(t_0) | {p\alpha'}}
	\nonumber \\
	&=
	\sum_{\nu} \braket{{p\alpha} | \Psi_{p\nu}(t)}\braket{ \Psi_{p\nu}(t_0) | {p\alpha'}}
	\nonumber \\
	&=
	\sum_{\nu} e^{-\im \epsilon_\nu (t-t_0)} \braket{{p\alpha} | \Phi_{p\nu}(t)}\braket{ \Phi_{p\nu}(t_0) | {p\alpha'}} \, .
\label{Eq:Utime}
\end{align}
In combination with the Fourier expansion of the Floquet states in Eq.~(\ref{Eq:State_Floquet_FS}), we obtain the full time dependence of any observable from the eigenvalue solution of Eq.~(\ref{Eq:FloquetHam}). This is particularly useful if we want to calculate integrals over time, as it is the case for the steady-state spectral functions considered below.

To shorten the notation below, we define the fermionic annihilation operators in
the Floquet basis as
\begin{align}
\fan{p\nu}(t_0)
=
\sum_{\alpha} \braket{ \Phi_{p\nu}(t_0) | {p\alpha}} \fan{p\alpha}(t_0) \, .
\end{align}

\subsection{Spectral functions}

Starting from the retarded and lesser Green's functions
\begin{gather}
\label{Eq:Gret_def}
 G^{\mathrm{ret}}_{p,\alpha \beta}(t,t')
	=
	- \im \Theta(t-t')
	\expv{ \left\{ \fan{p\alpha}(t), \fcr{p\beta}(t') \right\}} \, ,
\\
G^{<}_{p,\alpha \beta}(t,t')
	=
	\im \expv{\fcr{p\beta}(t') \, \fan{p\alpha}(t)} \, ,
\label{Eq:Gles_def}
\end{gather}
we can determine the spectral properties of the Floquet system.
In accordance with the solution of the interacting model at finite temperatures,
we use the Wigner coordinates
\begin{align}
\label{Eq:Wigner}
\tave = (t + t')/2 \, , \qquad
\trel = t - t' \, ,
\end{align}
to define the Fourier transform with respect to relative time as
\begin{align}
\nonumber
G_\mathrm{p,\alpha\alpha}(\tave, \omega)
	&=
	\int_{-\infty}^\infty d\trel \,
	e^{\im(\omega+\im \eta)\trel} 
	\\
	& \times G_{p,\alpha \alpha}(\tave+\trel/2,\tave-\trel/2) \, .
\end{align}
For the interacting model at finite temperatures, we find that the system reaches
a steady state where the local spectral functions
turn out to be positive semidefinite for any $\tave$ that is large enough.
For the zero-temperature case with a single phonon configuration,
the system does not decay but keeps oscillating forever.
It has been proved that the spectral function of the retarded Green's function is positive semidefinite
if one averages $\tave$ over the Floquet period $T$ \cite{PhysRevLett.122.130604}.
However, in our single-particle calculation for the lesser Green's function,
we need to average over time-dependent exponentials including the real-valued
quasienergies $\epsilon_\nu \in [-\Omega/2,\Omega/2)$ as follows:
\begin{align}
\lim_{T\to\infty} &\frac{1}{T} \int_{-T/2}^{T/2} d\tave \,
e^{-\im [\epsilon_\mu - \epsilon_\nu + (m-m')\Omega] \tave}
\nonumber \\
&=
\lim_{T\to\infty}
\mathrm{sinc} \{ [\epsilon_\mu - \epsilon_\nu + (m-m')\Omega] T/2 \}
\nonumber \\
&=
\delta_{\mu \nu} \, \delta_{mm'} \, .
\end{align}
Because $\mathrm{sinc}(0)=1$, $\mathrm{sinc}(\pm\infty) = 0$, and assuming that the $\epsilon_\nu$ are nondegenerate,
we obtain the Kronecker delta.
Hence, we define the spectral functions as follows:
\begin{align}
\bar{A}_{p\alpha}(\omega) &= -
\lim_{T\to\infty} \frac{1}{\pi T} \int_{-T/2}^{T/2} d\tave \,
\mathrm{Im} \, 
G^\mathrm{ret}_\mathrm{p,\alpha\alpha}(\tave,\omega) \, ,
\\
\bar{A}^{<}_{p\alpha}(\omega) &=
\lim_{T\to\infty} \frac{1}{2\pi T} \int_{-T/2}^{T/2} d\tave \,
\mathrm{Im} \, 
G^\mathrm{<}_\mathrm{p,\alpha\alpha}(\tave,\omega) \, .
\end{align}
This definition assumes that the initial time when the field is turned on fulfills $t_0\to - \infty$.
As discussed in more detail below, these definitions correspond to the steady-state spectra at $\kB T \to 0$.

In the following, we derive analytic expressions for the spectral functions. Because we are using the temporal gauge, the $p$ resolved spectral functions will be gauge dependent. We are mainly interested in the local spectra which are summed over all $p$ and therefore become gauge invariant again. To calculate momentum dependent observables, we have to substitute $p\to p-E \tave$ in the corresponding Green's functions before we perform the time average \cite{PhysRevB.44.3655}. We will discuss the necessary changes further below.

\subsubsection{Retarded Green's function}

We first calculate the retarded Green's function. Plugging Eqs.~(\ref{Eq:EOM})
and (\ref{Eq:Utime}) into Eq.~(\ref{Eq:Gret_def}), we obtain
\begin{align}
 &G^{\mathrm{ret}}_{p,\alpha \beta}(t,t')
	=
	- \im \Theta(t-t')
	\sum_{\mu \nu} 
	e^{-\im \epsilon_\mu (t-t_0)} e^{\im \epsilon_\nu (t'-t_0)} 
	\nonumber \\
	&\quad \times \braket{{p\alpha} | \Phi_{p\mu}(t)}
	\braket{\Phi_{p\nu}(t') | {p\beta} } 
	\expv{\{\fan{p\mu}(t_0), \fcr{p\nu}(t_0) \}} \, .
\end{align}
Because
$\{\fan{p\mu}(t_0), \fcr{p\nu}(t_0) \} = \delta_{\mu\nu}$,
we find
\begin{align}
 G^{\mathrm{ret}}_{p,\alpha \beta}(t,t')
	=
	- &\im \Theta(t-t')
	\sum_{\mu}
	e^{-\im \epsilon_\mu (t-t')}
	\nonumber \\
	&\times \braket{{p\alpha} | \Phi_{p\mu}(t)}
	 \braket{ \Phi_{p\mu}(t') | {p\beta}} \, .
\end{align}
Using the time evolution of the Floquet states, we have
\begin{align}
 G^{\mathrm{ret}}_{p,\alpha \beta}(t,t')
	=
	& - \im \Theta(t-t')
	\sum_{\mu} \sum_{mm'}
	e^{-\im \epsilon_\mu (t-t')}
	 e^{-\im m \Omega t}
	\nonumber \\
	&\times \, e^{ \im m' \Omega t'}
	 \braket{{p\alpha} | \phi_{p\mu}^{(m)}}
	 \braket{ \phi_{p\mu}^{(m')} | {p\beta}} \, .
\label{Eq:Gret_no_transf}
\end{align}
If we switch to the Wigner coordinates of Eq.~(\ref{Eq:Wigner}),
the time-dependent exponentials become
\begin{align}
e^{-\im [\epsilon_\mu + (m+m') \Omega/2] \trel} \,
e^{-\im (m-m') \Omega \tave } \, .
\end{align}
We can now do the Fourier transform in relative time and obtain
\begin{align}
\label{Eq:Gret_1}
 G^{\mathrm{ret}}_{p,\alpha \beta}(\tave, \omega)
=
	- 
	\sum_{\mu} \sum_{mm'}
	 & \frac{
	\braket{{p\alpha} | \phi_{p\mu}^{(m)}}
	 \braket{ \phi_{p\mu}^{(m')} | {p\beta}}
	}{\epsilon_\mu + (m+m') \Omega/2 - \omega - \im \eta}
	\nonumber \\
	&\times \,e^{-\im (m-m') \Omega \tave} \, .
\end{align}
Here, we explicitly see that we have to average over a period $T$ to obtain
a positive spectral function
\begin{align}
\bar{A}_{p\alpha}(\omega)
	=
	\sum_{\mu} \sum_{m}
	\absolute{\braket{{p\alpha} | \phi_{p\mu}^{(m)}}}^2
	\delta(\epsilon_\mu + m \Omega - \omega ) \, .
\end{align}

To show that the local spectral function is positive, we do not need to
average over $\tave$. We can just sum over $p$ and $\alpha$ in Eq.~(\ref{Eq:Gret_no_transf})
and use the symmetries in Eqs.~(\ref{Eq:State_p_transf}) and (\ref{Eq:State_subl_transf}) to obtain
\begin{align}
\frac{1}{L} \sum_{p\in[0,\pi)} &\sum_\alpha 
\braket{{p\alpha} | \phi_{p\mu}^{(m)}}
	 \braket{ \phi_{p\mu}^{(m')} | {p\alpha}}
\nonumber \\
&=
\frac{1}{L}
\sum_{p\in[0,\pi)} e^{-\im p (m-m')} \left( 1 + e^{-\im \pi (m-m')} \right)
\nonumber \\
&\quad\times \braket{{p=0,\alpha=1} | \phi_{p=0,\mu}^{(m)}}
	 \braket{ \phi_{p=0,\mu}^{(m')} | {p=0,\alpha=1}}
\nonumber \\
&=
\frac{1}{L}
\sum_{p\in[0,2\pi)} e^{-\im p (m-m')}
\nonumber \\
&\quad\times \braket{{p=0,\alpha=1} | \phi_{p=0,\mu}^{(m)}}
	 \braket{ \phi_{p=0,\mu}^{(m')} | {p=0,\alpha=1}}
\nonumber \\
&=
\delta_{mm'}
\absolute{\braket{{p=0,\alpha=1} | \phi_{p=0,\mu}^{(m)}}}^2 \, .
\label{Eq:Gret_psum}
\end{align}
Here, the average over $p$ has the same effect as the average over a period $T$.

\subsubsection{Lesser Green's function}

In the same way, we find that the lesser Green's function satisfies
\begin{align}
& -\im \, G^{<}_{p,\alpha \beta}(t,t')
	= \sum_{\mu \nu} 
	e^{-\im \epsilon_\mu (t-t_0)} e^{\im \epsilon_\nu (t'-t_0)} 
	\nonumber \\
	&\qquad\times \braket{{p\alpha} | \Phi_{p\mu}(t)}
	\braket{ \Phi_{p\nu}(t') | {p\beta}}
	\expv{\fcr{p\nu}(t_0) \fan{p\mu}(t_0)}  \, .
\label{Eq:Gles_start}
\end{align}
The dependence on the equal-time expectation value
at $t_0$ makes computations  more difficult.
Using the Fourier expansion of the Floquet states, we have
\begin{align}
&-\im \, G^{<}_{p,\alpha \beta}(t,t')
	= \sum_{\mu \nu} 
	\sum_{mm'}
	e^{-\im (\epsilon_\mu + m\Omega) t} e^{\im (\epsilon_\nu + m' \Omega) t'} e^{\im (\epsilon_\mu -\epsilon_\nu) t_0}
	\nonumber \\
	&\qquad \quad \times \braket{{p\alpha} | \phi_{p\mu}^{(m)}}
	 \braket{ \phi_{p\nu}^{(m')} | {p\beta}}
	 \expv{\fcr{p\nu}(t_0) \fan{p\mu}(t_0)} \, .
\end{align}
If we switch to Wigner coordinates, the time-dependent exponentials become
\begin{align}
e^{-\im [\epsilon_\mu - \epsilon_\nu + (m-m')\Omega] \tave}
e^{-\im [\epsilon_\mu + \epsilon_\nu + (m+m')\Omega] \trel/2}
e^{\im (\epsilon_\mu -\epsilon_\nu) t_0} \, .
\end{align}
As discussed above, we have to average $\tave$ over all times to obtain a positive weight.
We find that
\begin{align}
- \im \, \overline{G}^{<}_{p,\alpha \beta}(\omega)
= &2\pi \sum_{\mu} 
	\sum_{m}
 	\braket{{p\alpha} | \phi_{p\mu}^{(m)}}
	 \braket{ \phi_{p\mu}^{(m)} | {p\beta} }
	\nonumber \\
	&\times \expv{\fcr{p\mu}(t_0) \fan{p\mu}(t_0)}
	\delta( \epsilon_\mu  + m \Omega - \omega ).
\end{align}
From this, we can see that the momentum resolved spectrum
$A^{<}_{p\alpha}(\omega)$
(and therefore also the local spectrum)
is always positive semidefinite, \ie,
\begin{align}
\bar{A}^<_{p\alpha}(\omega)
	=
	\sum_{\mu m} 
	\big\|
	& \fan{p\mu}(t_0) \ket{\mathrm{GS}}
	\big\| ^2
	  \absolute{ \braket{{p\alpha} | \phi_{p\mu}^{(m)}} }^2
	  \nonumber \\
	&\qquad \times \delta( \epsilon_\mu  + m \Omega - \omega ) \, .
\end{align}
For simplicity of notation, we assumed that the expectation value is with respect
to the ground state $\ket{\mathrm{GS}}$.

In contrast to the retarded Green's function, there is no obvious way to show that the spectrum of the local lesser Green's function is positive without also performing a time average. The nontrivial momentum dependence of the initial state makes this calculation much more difficult. After performing the time average, it becomes positive semidefinite as it must, since it is just a sum over all momenta of the positive semidefinite $A_{p\alpha}^<(\omega)$.

\subsubsection{Distribution function}

We can now calculate the nonequilibrium distribution function as the ratio
of lesser and retarded spectral functions, \ie,
\begin{align}
\bar{f}_{\infty,p\alpha}(\omega) =
\frac{  \bar{A}^<_{p\alpha} (\omega) }
{\bar{A}_{p\alpha}(\omega)} \, .
\end{align}
For our noninteracting model at $\kB T = 0$ we find
\begin{align}
\bar{f}_{\infty,p\alpha}(\epsilon_\mu + m \Omega)
=
  \expv{\fcr{p\mu}(t_0) \fan{p\mu}(t_0)} \, .
\end{align}
Interestingly, the distribution function averaged over all time corresponds
to the average occupation of a Floquet state
at the initial time.
Therefore, the distribution function for a Floquet level is determined by the overlap of the corresponding Floquet state with the initial state. However, this $p$ resolved ratio is not gauge invariant. To this end,
we calculate the ratio of the local spectral functions and see that
\begin{align}
\bar{f}_\infty(\epsilon_\mu + m \Omega)
=
\frac{2}{L} \sum_{p\in[0,\pi)} \expv{\fcr{p\mu}(t_0) \fan{p\mu}(t_0)} \, .
\end{align}
Here, we used Eq.~(\ref{Eq:State_p_transf}) to show that $| \braket{{p\alpha} | \phi_{p\mu}^{(m)}} |^2$
is independent of $p$ and therefore drops out of the ratio.
Remarkably, the local ratio $\bar{f}_\infty$ fully determines the gauge-invariant steady-state observables, even the momentum-dependent ones. This will become clear below.

\subsection{Steady-state observables}

Although the time-dependent observables at zero temperature do not decay towards a steady state, we can define the time average
\begin{align}
\label{Eq:TimeAvSS}
\expv{\hat{O}}_\infty
=
\lim_{T\to\infty}
\frac{1}{T}
 \int_{t_0}^{t_0+T} dt \expv{\hat{O}(t)} \, ,
\end{align}
which is consistent with the steady-state value in the interacting model.
Here, $t_0$ is the time when the field is turned on.
We will see below that this definition reproduces compatible results.

\subsubsection{Gauge-invariant momentum distribution function}

We want to calculate the momentum-distribution function in the steady state.
For this, we have to consider
the gauge-invariant form \cite{PhysRevB.44.3655}
\begin{align}
n(k,t)
=
-\im \, G^<_{p-Et,\alpha\alpha}(t,t)
\end{align}
where $k=p+(\alpha-1)\pi$. Starting from the lesser Green's function defined in Eq.~(\ref{Eq:Gles_start}), we have to properly incorporate the time-dependent momentum shifts.
First, we simplify the matrix elements
$\braket{{p-Et,\alpha} | \Phi_{p-Et,\mu}(t)}$
using the Fourier expansion in Eq.~(\ref{Eq:State_Floquet_FS}) with $\Omega=E$ and then using the momentum translation symmetry in Eq.~(\ref{Eq:State_p_transf}). We find that
\begin{align}
    & \braket{{p-Et,\alpha} | \Phi_{p-Et,\mu}(t)}
    =
    \sum_m e^{-\im m \Omega t}
    \braket{{p-Et,\alpha} | \phi_{p-Et,\mu}^{(m)}}
    \nonumber\\
    &\qquad=
    \sum_m e^{-\im m \Omega t}
    e^{-\im m (p - E t)}
    \braket{{p=0,\alpha} | \phi_{p=0,\mu}^{(m)}}
    \nonumber\\
    &\qquad=
    \braket{{p\alpha} | \Phi_{p\mu}(t=0)} \, ,
\end{align}
where the time dependence has dropped out due to the gauge transformation.
Second, we observe that the expectation value
$\expv{\fcr{p-Et,\nu}(t_0) \fan{p-Et,\mu}(t_0)}$
is a periodic function in time with period $T=2\pi/E$,
so that it can be expanded in a Fourier series.
Its time dependence is given by $e^{-\im \tilde{m}\Omega t}$
and together with the factor $e^{-\im (\epsilon_\mu -\epsilon_\nu) t}$ from Eq.~(\ref{Eq:Gles_start}), the time average in Eq.~(\ref{Eq:TimeAvSS}) leads to $\delta_{\mu \nu} \, \delta_{\tilde{m}0}$.
The zeroth element of the expectation value is given by
\begin{align}
    &\frac{1}{T} \int_0^T dt \expv{\fcr{p-Et,\mu}(t_0) \fan{p-Et,\mu}(t_0)}
    \nonumber \\
    &\qquad \cong \frac{2}{L} \sum_p
    \expv{\fcr{p\mu}(t_0) \fan{p\mu}(t_0)}
    = \bar{f}_\infty(\epsilon_\mu) \, .
\end{align}
Because in the thermodynamic limit the time average over a period is equivalent to an average over all momenta, the expectation value reduces to the local distribution function $\bar{f}_\infty(\epsilon_\mu)$. With this, the gauge-invariant
momentum distribution function becomes
\begin{align}
    \bar{n}_\infty(k)
    =
    \sum_\mu
    \absolute{\braket{{p\alpha} | \Phi_{p\mu}(t=0)}}^2
    \bar{f}_\infty(\epsilon_\mu) \, .
\end{align}
In particular, if $\bar{f}_\infty(\epsilon_\mu) = 1/2$ for both Floquet levels $\mu$, we can use the completeness of the Floquet states to show that $\bar{n}_\infty(k) = 1/2$ for all $k$,
as we would expect for an infinite-temperature state.

\subsubsection{Total energy}

To determine the total energy of the steady state, we do not need to use the gauge-invariant Green's function because we sum over all $p$.
Again, we can easily calculate the average over time using Floquet theory.
For
single-particle observables at equal times, we expand
\begin{align}
\label{Eq:obs_ss_eq}
\hat{O}(t)
=
\sum_p \sum_{\alpha\beta} O_{p,\alpha \beta}(t)
\left[-\im \, G^<_{p,\beta \alpha}(t,t) \right] \, .
\end{align}
We can directly calculate the electron-phonon energy, because its matrix element
has no time dependence. We have that
\begin{align}
E^\infty_\mathrm{eph}
&=
\Delta \sum_{p\alpha} \sum_{\mu m}
\braket{{p\alpha} | \phi_{p\mu}^{(m)}}
 \braket{ \phi_{p\mu}^{(m)} | {p\bar{\alpha}}} 
   \expv{\fcr{p\mu}(t_0) \fan{p\mu}(t_0)} \, .
\end{align}
For the kinetic energy, we must include the time dependence of the electric
field. Therefore, we use the Fourier series of the matrix element to obtain
\begin{align}
E_\mathrm{kin}^\infty
&=
\sum_{p\alpha} \sum_{s=\pm1}
{\mathcal{H}}_{p,\alpha\alpha}^{(s)}
\sum_{\mu m}  
	 \braket{{p\alpha} | \phi_{p\mu}^{(m)}}
	\nonumber \\
	&\qquad\times \braket{ \phi_{p\mu}^{(m+s)} | {p\alpha}} 
	 \expv{\fcr{p\mu}(t_0) \fan{p\mu}(t_0)} \, .
\end{align}

A better understanding of the steady-state energies can be obtained by starting from the gauge-invariant form. If we substitute $p\to p-Et$ in Eq.~(\ref{Eq:obs_ss_eq}), we find that $\mathcal{H}_{p-Et,\alpha\beta}(t) = \mathcal{H}_{p,\alpha\beta}(t=0)$ loses its time dependence.
Therefore, the time average only applies to the lesser Green's function. As before we can derive
\begin{align}
    -\im \, G^<_{p-Et,\beta\alpha}(t,t)
    =
    \sum_\mu
    &\braket{{p\beta} | \Phi_{p\mu}(t=0)} 
    \nonumber \\
    \times &\braket{\Phi_{p\mu}(t=0) | {p\alpha} }
    \bar{f}_\infty(\epsilon_\mu) \, .
\end{align}
If $\bar{f}_\infty(\epsilon_\mu) =1/2$ for both Floquet levels,
we can use the completeness of the Floquet states to obtain the matrix element $\braket{p\beta | p \alpha } = \delta_{\alpha\beta}$.
Then, we find that $E^\infty_\mathrm{eph}=0$ because $\mathcal{H}_{p,\alpha\beta}(t=0)$
has only off-diagonal entries.
We also find that $E^\infty_\mathrm{kin}=0$ because the diagonal elements cancel each other.
Again, these results are consistent with an infinite-temperature state.

\subsubsection{Gauge-invariant spectral functions}

Finally, we want to discuss how the spectral functions change
if we start from the gauge-invariant Green's functions. For the retarded Green's function, the substitution $p\to \bar{p}=p-E\tave$ will only eliminate the factor $e^{-\im(m-m')\Omega \tave}$ in Eq.~(\ref{Eq:Gret_1}) and we obtain 
\begin{align}
  \bar{A}_{\bar{p}\alpha}(\omega)
	&=
	\sum_{\mu} \sum_{mm'}
	\braket{{p\alpha} | \phi_{p\mu}^{(m)}}
	\braket{ \phi_{p\mu}^{(m')} | {p\alpha} }
	\nonumber \\
	&\qquad\qquad\times \delta[\epsilon_\mu + (m+m') \Omega/2 - \omega ] \, . 
\end{align}
For the lesser Green's function, we have to average over $\tave$ to obtain
\begin{align}
  \bar{A}^<_{\bar{p}\alpha}(\omega)
= &2\pi \sum_{\mu}
     \bar{f}_\infty(\epsilon_\mu)
	\sum_{mm'}
 	\braket{{p\alpha} | \phi_{p\mu}^{(m)}}
	 \braket{ \phi_{p\mu}^{(m')} | {p\alpha} }
	\nonumber \\
	&\qquad \qquad\times
	\delta[ \epsilon_\mu  + (m+m') \Omega/2 - \omega ] \, .
\end{align}
In the gauge-invariant form, the ratio of the momentum-resolved
spectral functions is just given by the local ratio $\bar{f}_\infty(\epsilon_\mu)$. Although we cannot prove positive-definiteness for either of the spectra, their ratio is given by a positive function. 
To obtain positive spectral functions, we still have to sum over all momenta as in Eq.~(\ref{Eq:Gret_psum}) in order to reproduce our previous results.

\section{Additional results}

\subsection{Zero temperature}

\begin{figure}[tbp]
  \includegraphics[width=\linewidth]{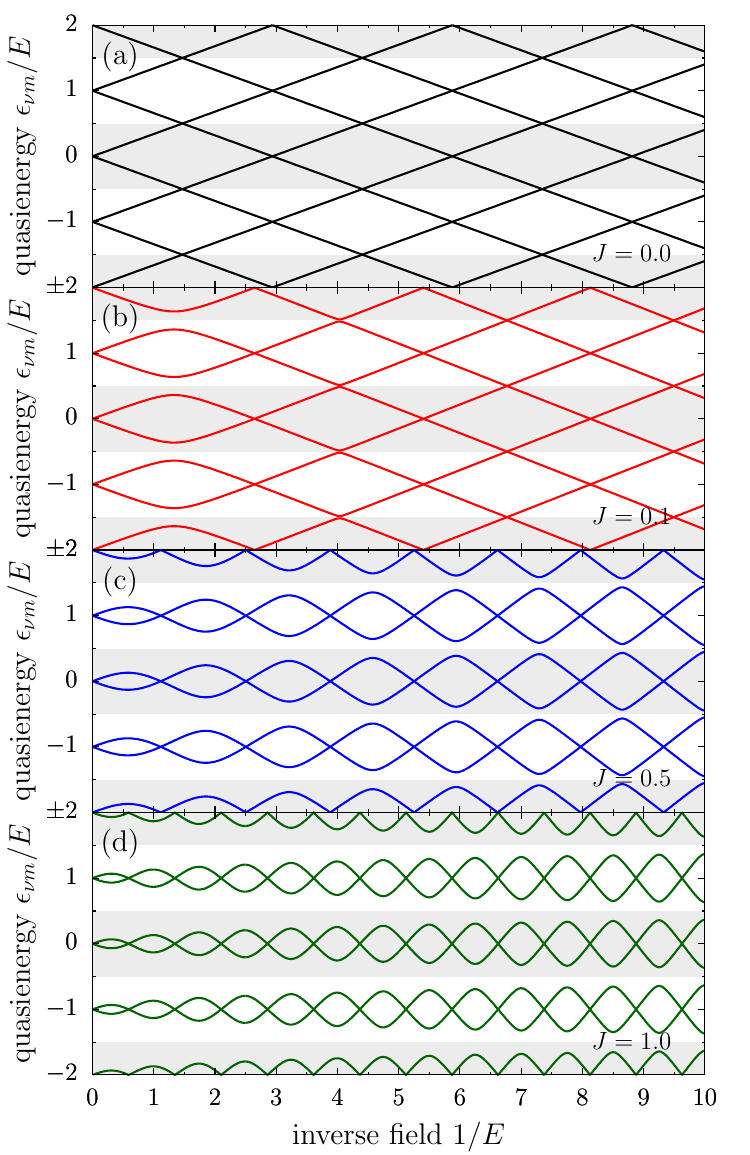}
  \caption{\label{fig:energygap}%
Floquet quasienergies for $\Delta=0.3404$ as a function of inverse electric field
for (a) $J=0.0$, (b) $J=0.1$, (c) $J=0.5$, and (d) $J=1.0$. The hopping amplitude $J$
leads to avoided level crossings at the Floquet zone boundaries.
  }
\end{figure}
As we have seen in the previous section, the Holstein model
can be solved efficiently at zero temperature using Floquet theory. Although
the system never reaches a steady state with zero current and constant energy at $\kB T = 0$
(as occurs for all $\kB T > 0$), we still obtain important insights from the $\kB T=0$ solution
into the nature of the steady state and the heating process at $\kB T > 0$. In the following,
we expand on our discussion of the $\kB T = 0$ case presented in the main article.

\begin{figure}[t]
  \includegraphics[width=\linewidth]{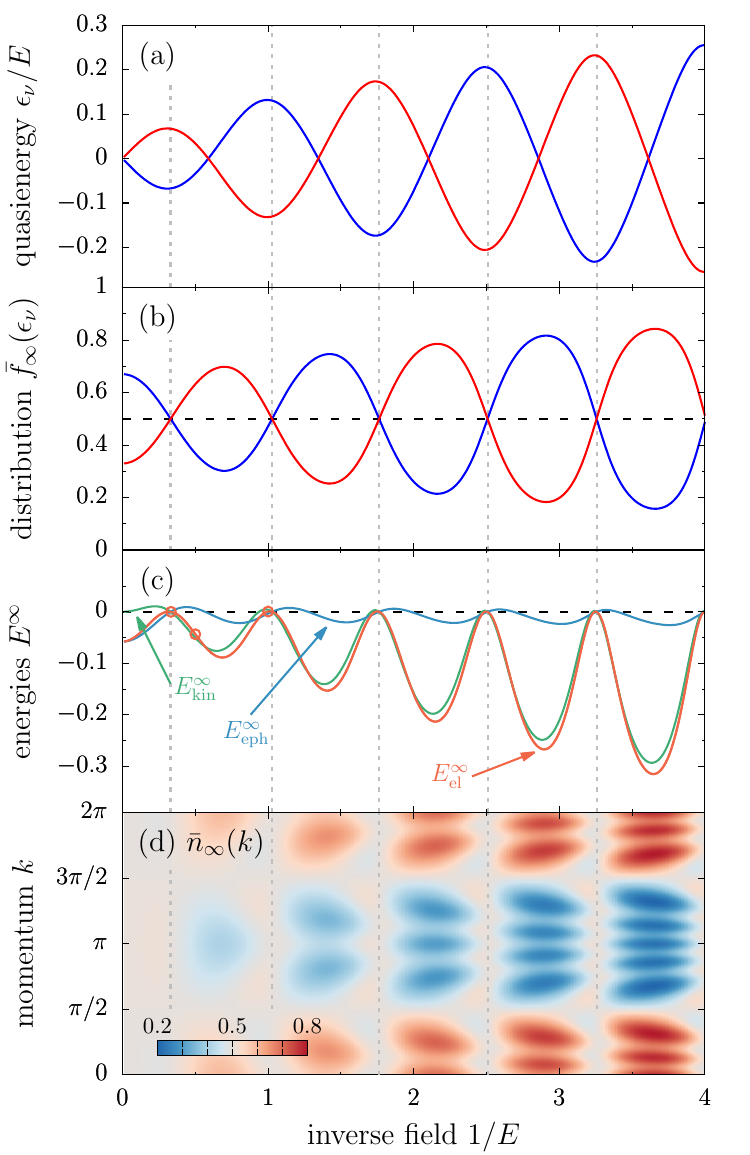}
  \caption{\label{fig:Floquet}%
Solution of the Holstein model at $\kB T = 0$ as a function of inverse
electric field. (a) Floquet quasienergies within the first Floquet zone and
(b) the corresponding distribution function $\bar{f}_\infty(\epsilon_\nu)$.
(c) Time-averaged energies.
(d) Gauge-invariant momentum distribution function.
The black dashed lines in (b) and (c) correspond to the infinite-temperature solution.
Vertical lines mark the electric-field values where avoided level crossings occur.
Open circles in (c) correspond to the exact ground-state energies of the fields considered
in Fig.~\ref{fig:energies}.
Here, $\Delta=0.3404$ and $J=1$.
  }
\end{figure}

The Floquet energy spectrum of our two-band model consists of two interpenetrating
Wannier-Stark ladders. Figure~\ref{fig:energygap}
illustrates the gap opening at the Floquet zone boundaries
for different hopping amplitudes $J$ and fixed $\Delta=0.3404$. For $J\to 0$, the two ladders are independent
of each other and the exact energies are
$\epsilon_{\nu m}  = \pm \Delta + m E$. The corresponding energy spectrum
in Fig.~\ref{fig:energygap}(a) shows level crossings both at the Floquet zone  boundary and center.
With increasing $J$, the spectrum opens up a gap at the Floquet zone boundary
that is largest for strong fields, as shown in Figs.~\ref{fig:energygap}(b)--(d). Moreover,
the avoided level crossings shift towards stronger fields with increasing $J$.
On the other side, the level crossings at the Floquet zone center remain.

In Fig.~\ref{fig:Floquet}, we compare the occurrence of level crossings and avoided crossings
in the energy spectrum to the system's ability to distribute the energy absorbed from the electric field.
Figure~\ref{fig:Floquet}(a) shows the two quasienergy levels $\epsilon_\nu$ in the first Floquet zone
as a function of $1/E$.
The corresponding distribution functions
$\bar{f}_\infty(\epsilon_\nu)$ in Fig.~\ref{fig:Floquet}(b) give the time-averaged probability of finding an electron in one of the levels. At a set of exceptional points
we find $\bar{f}_\infty(\epsilon_\nu)=1/2$ for both levels, which is the analog to an infinite-temperature state. Away from these points, either the lower or the upper Floquet level
has a higher occupation which corresponds to an effective positive or negative temperature.
Moreover, the time-averaged electronic energy in Fig.~\ref{fig:Floquet}(c)
reaches
its infinite-temperature limit $E_\mathrm{el} = 0$ at these avoided level crossings.
Furthermore, at these resonances, the system efficiently redistributes
the energy absorbed from the electric field.
In contrast, near the level crossings at the Floquet zone center, the average energy
is much lower. We will demonstrate below that the time-averaged energies of
the non-decaying system at $\kB T = 0$ correctly predicts the $\kB T \to 0$ limit
of the true steady state at finite initial temperatures.
Finally, the gauge-invariant momentum distribution function $\bar{n}_\infty(k)$ in Fig.~\ref{fig:Floquet}(d)
becomes completely flat when $\bar{f}_\infty(\epsilon_\mu) =1/2$, as proved above.
Away from the avoided level crossings, $\bar{n}_\infty(k)$ increasingly gains more features with each resonance that is crossed.
A similar structure had been observed for the time-dependent oscillations of the density matrix \cite{PhysRevLett.74.1831}.
From Fig.~\ref{fig:Floquet}(d), it seems that $\bar{n}_\infty(k)<1/2$ for all $\pi/2 < k < 3\pi/2$, as in the initial state.
\begin{figure}[t]
  \includegraphics[width=\linewidth]{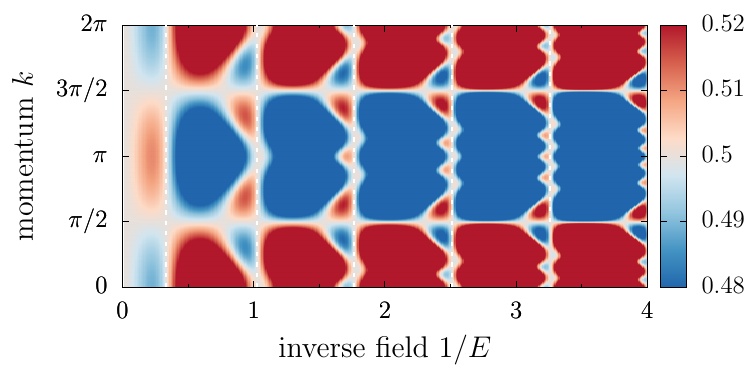}
  \caption{\label{fig:nk}%
Close-up of the gauge-invariant momentum distribution function shown in Fig.~\ref{fig:Floquet}(d).
For better visibility of the delocalization features, we restrict the color range to $\bar{n}_\infty(k) \in [0.48,0.52]$.
Vertical lines mark the electric-field values where avoided level crossings occur.
Here, $\Delta=0.3404$ and $J=1$.
  }
\end{figure}
For better visibility of the detailed structure near $\bar{n}_\infty(k)=1/2$,
we show the same data again in Fig.~\ref{fig:nk} but for a smaller range of $\bar{n}_\infty(k)$. Close to the resonances, we also find regimes where the sign structure of $\bar{n}_\infty(k)$ is reversed.

\begin{figure}
  \includegraphics[width=\linewidth]{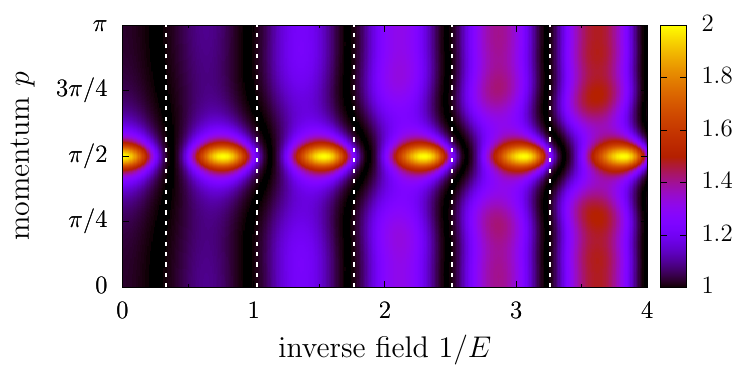}
  \caption{\label{fig:IPR}%
Inverse participation ratio, as defined in Eq.~(\ref{eq:IPR}), as a function of inverse field and momentum.
Vertical lines mark the electric-field values where avoided level crossings occur.
Here, $\Delta=0.3404$ and $J=1$.
  }
\end{figure}

The avoided level crossings of the field-driven two-band model have been associated
with resonance-induced delocalization in coupled Wannier-Stark ladders, both
theoretically \cite{Leo_1989} and experimentally \cite{PhysRevLett.65.2720}.
We can gain further insight into the localization properties
of our Floquet system
from the
inverse participation ratio. We define
\begin{align}
\label{eq:IPR}
\mathrm{IPR}(p)
=
\sum_{s\mu}   \absolute{ \braket{ps|\Phi_{p\mu}(t=0)}}^4
\end{align}
from the overlap of the Floquet states at $t=0$ with the energy
eigenstates $\ket{ps}$ of the system at equilibrium.
We have $\mathrm{IPR}(p) = 2$ if the Floquet states
are perfectly localized in the energy eigenbasis, \ie, they coincide with one of
the two equilibrium eigenstates. On the other hand, $\mathrm{IPR}(p) = 1$
for perfectly delocalized Floquet states that are an equal superposition
of the equilibrium eigenstates.
Figure~\ref{fig:IPR} shows $\mathrm{IPR}(p)$ as a function of $1/E$.
Indeed, the avoided level crossings 
appear at the field strengths where the equilibrium eigenstates transform into both Floquet states with equal weight,
but the delocalization is not perfect, since the
minimum of $\mathrm{IPR}(p)$ slightly
depends on $p$.
Moreover,
Wannier-Stark localization
is strongest between the avoided crossings and near $p=\pi/2$, where the
gap of the equilibrium system is smallest.
We have seen that $\mathrm{IPR}(p)$ gives us some insight into the delocalization properties, but one has to be careful with far-reaching conclusions.
$\mathrm{IPR}(p)$ only tells us how the initial eigenstates transform into the Floquet states at $t=0$. While the latter govern the time evolution by a period $T$, they do not tell us how states are occupied at intermediate times. To obtain properly-defined steady-state averages, we need to know the occupation at all times.
In addition, the issue of gauge invariance would occur again if we want to get access to the physical momentum $k$.

\begin{figure}
  \includegraphics[width=\linewidth]{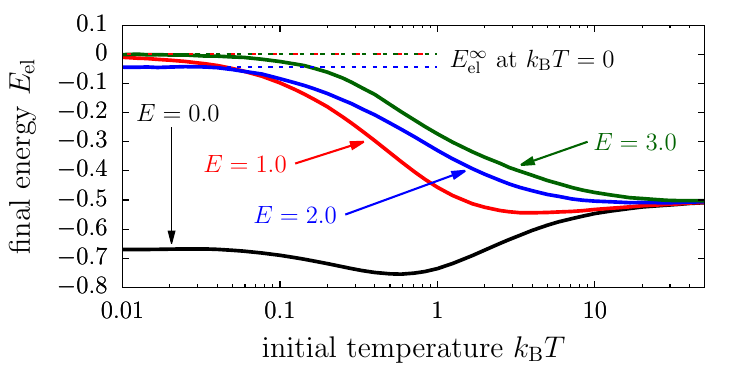}
  \caption{\label{fig:energies}%
Electronic energy of the steady state as a function of initial temperature for different electric fields $E$. The dashed lines illustrate the time-averaged energies $E^\infty_\mathrm{el}$ at $\kB T=0$.
Here $L=42$, $\lambda=0.5$.
  }
\end{figure}

\subsection{Finite temperatures}

\begin{figure}
  \includegraphics[width=\linewidth]{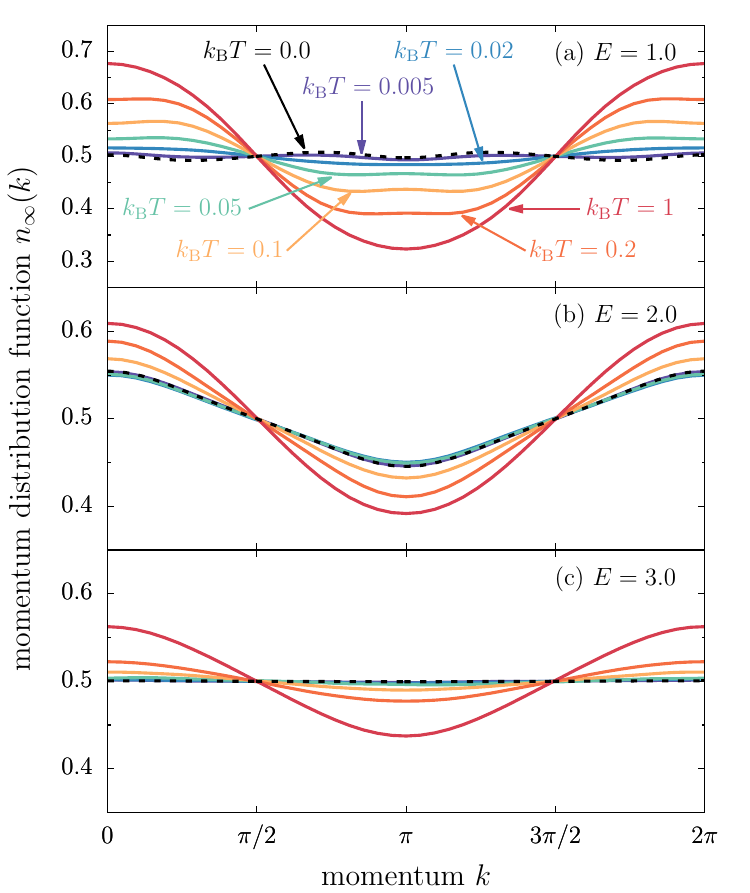}
  \caption{\label{fig:nktemp}%
Momentum distribution function of the steady state for different electric fields $E$ and initial temperatures $\kB T$.
Here $L=42$, $\lambda=0.5$.
  }
\end{figure}

After having expanded on the zero-temperature solution, we also want to extend our discussion
of the interacting system at $\kB T > 0$. Figure~\ref{fig:energies} shows
the electronic energy of the steady state as a function of the initial temperature.
We compare different electric fields to the equilibrium solution at $E=0$. Note
that the electronic energy is allowed to decrease with increasing $\kB T$ because the loss
is compensated by an increasing phonon potential energy that is not considered here; note that static phonons cannot absorb energy from the electrons.
As already discussed in our main article, the heating of the system is strongest at low initial
temperatures. In particular, the time-averaged energies at $\kB T = 0$---indicated
by the open circles in Fig.~\ref{fig:Floquet}(c)---perfectly predict the low-temperature
properties of the steady state. Therefore, the heating behavior at low $\kB T$
will show strong oscillations as a function of the electric-field strength,
as shown in Fig.~\ref{fig:Floquet}(c). In contrast, the high-temperature regime
only shows weak heating effects that increase with $E$. This is consistent
with the effect of an electric field on a strongly Anderson-localized system
\cite{Prigodin1980, PhysRevB.35.8929}.
While Anderson localization predominantly occurs at high $\kB T$, a crossover
towards Wannier-Stark localization occurs when we lower the temperature, as
field effects can overcome the decreasing phonon disorder.

\begin{figure}
  \includegraphics[width=\linewidth]{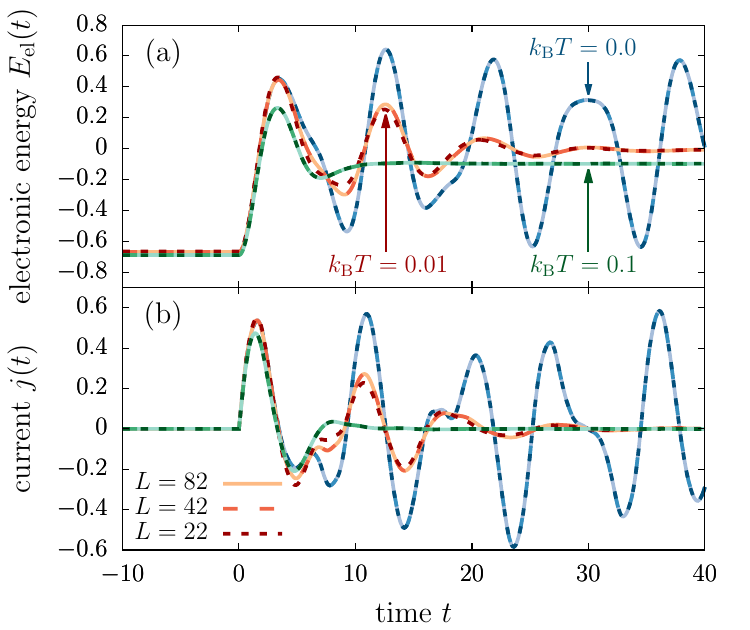}
  \caption{\label{fig:FS}%
Finite-size analysis of (a) the electronic energy and (b) the current for different initial
temperatures. Here, $\lambda=0.5$ and $E=1.0$.
  }
\end{figure}

Figure~\ref{fig:nktemp} shows the momentum distribution function of the steady state for different initial temperatures and for different electric fields. Again, we confirm that the finite-temperature results approach the time-averaged Floquet solution for $\kB T \to 0$.
With increasing $\kB T$, the variance of $n_\infty(k)$ increases up to $\kB T \approx 1$. This is a result of the phonon disorder which inhibits heating effects. At even higher temperatures, $n_\infty(k)$ will become flatter again, as expected for the infinite-temperature solution in equilibrium (not shown).
For $E=1$ [Fig.~\ref{fig:nktemp}(a)] we find that the small oscillations of the $\kB T=0$
solution survive up to $\kB T \approx 0.2$. When the field is tuned to a zero-temperature resonance [Fig.~\ref{fig:nktemp}(c)], $n_\infty(k)$ remains almost flat up to rather high temperatures of $\kB T\approx0.1$. These observations suggest that the Floquet physics remains relevant as long as the charge-density-wave correlations survive.
In the high-temperature regime, an increasing electric field drives the system closer to an infinite-temperature state, as discussed in the main text and for the electronic energies in Fig.~\ref{fig:energies}.

Finally, Fig.~\ref{fig:FS} provides a
finite-size analysis of the electronic energy and the current
as a function of time. We find that finite-size effects
are generically small when a constant  field is applied to the 1D Holstein model.
Lattice sizes of $L=42$ are sufficient for convergence within the size of the
linewidth. Moreover, finite-size effects do not seem to increase with time, which allows
us to evolve our system towards large average times $\tave$.


\begin{thebibliography}{38}%
\makeatletter
\providecommand \@ifxundefined [1]{%
 \@ifx{#1\undefined}
}%
\providecommand \@ifnum [1]{%
 \ifnum #1\expandafter \@firstoftwo
 \else \expandafter \@secondoftwo
 \fi
}%
\providecommand \@ifx [1]{%
 \ifx #1\expandafter \@firstoftwo
 \else \expandafter \@secondoftwo
 \fi
}%
\providecommand \natexlab [1]{#1}%
\providecommand \enquote  [1]{``#1''}%
\providecommand \bibnamefont  [1]{#1}%
\providecommand \bibfnamefont [1]{#1}%
\providecommand \citenamefont [1]{#1}%
\providecommand \href@noop [0]{\@secondoftwo}%
\providecommand \href [0]{\begingroup \@sanitize@url \@href}%
\providecommand \@href[1]{\@@startlink{#1}\@@href}%
\providecommand \@@href[1]{\endgroup#1\@@endlink}%
\providecommand \@sanitize@url [0]{\catcode `\\12\catcode `\$12\catcode
  `\&12\catcode `\#12\catcode `\^12\catcode `\_12\catcode `\%12\relax}%
\providecommand \@@startlink[1]{}%
\providecommand \@@endlink[0]{}%
\providecommand \url  [0]{\begingroup\@sanitize@url \@url }%
\providecommand \@url [1]{\endgroup\@href {#1}{\urlprefix }}%
\providecommand \urlprefix  [0]{URL }%
\providecommand \Eprint [0]{\href }%
\providecommand \doibase [0]{https://doi.org/}%
\providecommand \selectlanguage [0]{\@gobble}%
\providecommand \bibinfo  [0]{\@secondoftwo}%
\providecommand \bibfield  [0]{\@secondoftwo}%
\providecommand \translation [1]{[#1]}%
\providecommand \BibitemOpen [0]{}%
\providecommand \bibitemStop [0]{}%
\providecommand \bibitemNoStop [0]{.\EOS\space}%
\providecommand \EOS [0]{\spacefactor3000\relax}%
\providecommand \BibitemShut  [1]{\csname bibitem#1\endcsname}%
\let\auto@bib@innerbib\@empty
\bibitem [{\citenamefont {{Basov}}\ \emph {et~al.}(2017)\citenamefont
  {{Basov}}, \citenamefont {{Averitt}},\ and\ \citenamefont
  {{Hsieh}}}]{2017NatMa..16.1077B}%
  \BibitemOpen
  \bibfield  {author} {\bibinfo {author} {\bibfnamefont {D.~N.}\ \bibnamefont
  {{Basov}}}, \bibinfo {author} {\bibfnamefont {R.~D.}\ \bibnamefont
  {{Averitt}}},\ and\ \bibinfo {author} {\bibfnamefont {D.}~\bibnamefont
  {{Hsieh}}},\ }\bibfield  {title} {\bibinfo {title} {{Towards properties on
  demand in quantum materials}},\ }\href {https://doi.org/10.1038/nmat5017}
  {\bibfield  {journal} {\bibinfo  {journal} {Nature Materials}\ }\textbf
  {\bibinfo {volume} {16}},\ \bibinfo {pages} {1077} (\bibinfo {year}
  {2017})}\BibitemShut {NoStop}%
\bibitem [{\citenamefont {{Oka}}\ and\ \citenamefont
  {{Kitamura}}(2019)}]{2019ARCMP..10..387O}%
  \BibitemOpen
  \bibfield  {author} {\bibinfo {author} {\bibfnamefont {T.}~\bibnamefont
  {{Oka}}}\ and\ \bibinfo {author} {\bibfnamefont {S.}~\bibnamefont
  {{Kitamura}}},\ }\bibfield  {title} {\bibinfo {title} {{Floquet Engineering
  of Quantum Materials}},\ }\href
  {https://doi.org/10.1146/annurev-conmatphys-031218-013423} {\bibfield
  {journal} {\bibinfo  {journal} {Annual Review of Condensed Matter Physics}\
  }\textbf {\bibinfo {volume} {10}},\ \bibinfo {pages} {387} (\bibinfo {year}
  {2019})}\BibitemShut {NoStop}%
\bibitem [{\citenamefont {{Rudner}}\ and\ \citenamefont
  {{Lindner}}(2020{\natexlab{a}})}]{2020NatRP...2..229R}%
  \BibitemOpen
  \bibfield  {author} {\bibinfo {author} {\bibfnamefont {M.~S.}\ \bibnamefont
  {{Rudner}}}\ and\ \bibinfo {author} {\bibfnamefont {N.~H.}\ \bibnamefont
  {{Lindner}}},\ }\bibfield  {title} {\bibinfo {title} {{Band structure
  engineering and non-equilibrium dynamics in Floquet topological
  insulators}},\ }\href{https://doi.org/10.1038/s42254-020-0170-z} {\bibfield  {journal} {\bibinfo  {journal}
  {Nature Reviews Physics}\ }\textbf {\bibinfo {volume} {2}},\ \bibinfo {pages}
  {229} (\bibinfo {year} {2020}{\natexlab{a}})}\BibitemShut {NoStop}%
\bibitem [{\citenamefont {Khemani}\ \emph {et~al.}(2016)\citenamefont
  {Khemani}, \citenamefont {Lazarides}, \citenamefont {Moessner},\ and\
  \citenamefont {Sondhi}}]{PhysRevLett.116.250401}%
  \BibitemOpen
  \bibfield  {author} {\bibinfo {author} {\bibfnamefont {V.}~\bibnamefont
  {Khemani}}, \bibinfo {author} {\bibfnamefont {A.}~\bibnamefont {Lazarides}},
  \bibinfo {author} {\bibfnamefont {R.}~\bibnamefont {Moessner}},\ and\
  \bibinfo {author} {\bibfnamefont {S.~L.}\ \bibnamefont {Sondhi}},\ }\bibfield
   {title} {\bibinfo {title} {Phase structure of driven quantum systems},\
  }\href {https://doi.org/10.1103/PhysRevLett.116.250401} {\bibfield  {journal}
  {\bibinfo  {journal} {Phys. Rev. Lett.}\ }\textbf {\bibinfo {volume} {116}},\
  \bibinfo {pages} {250401} (\bibinfo {year} {2016})}\BibitemShut {NoStop}%
\bibitem [{\citenamefont {{Sacha}}\ and\ \citenamefont
  {{Zakrzewski}}(2018)}]{2018RPPh...81a6401S}%
  \BibitemOpen
  \bibfield  {author} {\bibinfo {author} {\bibfnamefont {K.}~\bibnamefont
  {{Sacha}}}\ and\ \bibinfo {author} {\bibfnamefont {J.}~\bibnamefont
  {{Zakrzewski}}},\ }\bibfield  {title} {\bibinfo {title} {{Time crystals: a
  review}},\ }\href {https://doi.org/10.1088/1361-6633/aa8b38} {\bibfield
  {journal} {\bibinfo  {journal} {Reports on Progress in Physics}\ }\textbf
  {\bibinfo {volume} {81}},\ \bibinfo {eid} {016401} (\bibinfo {year}
  {2018})}\BibitemShut {NoStop}%
\bibitem [{\citenamefont {{Khemani}}\ \emph {et~al.}(2019)\citenamefont
  {{Khemani}}, \citenamefont {{Moessner}},\ and\ \citenamefont
  {{Sondhi}}}]{2019arXiv191010745K}%
  \BibitemOpen
  \bibfield  {author} {\bibinfo {author} {\bibfnamefont {V.}~\bibnamefont
  {{Khemani}}}, \bibinfo {author} {\bibfnamefont {R.}~\bibnamefont
  {{Moessner}}},\ and\ \bibinfo {author} {\bibfnamefont {S.~L.}\ \bibnamefont
  {{Sondhi}}},\ }\bibfield  {title} {\bibinfo {title} {{A Brief History of Time
  Crystals}},\ }\href{https://doi.org/10.48550/arXiv.1910.10745} {\bibfield  {journal} {\bibinfo  {journal}
  {arXiv:1910.10745}} (\bibinfo {year} {2019})}\BibitemShut {NoStop}%
\bibitem [{\citenamefont {D'Alessio}\ and\ \citenamefont
  {Rigol}(2014)}]{PhysRevX.4.041048}%
  \BibitemOpen
  \bibfield  {author} {\bibinfo {author} {\bibfnamefont {L.}~\bibnamefont
  {D'Alessio}}\ and\ \bibinfo {author} {\bibfnamefont {M.}~\bibnamefont
  {Rigol}},\ }\bibfield  {title} {\bibinfo {title} {{Long-time Behavior of
  Isolated Periodically Driven Interacting Lattice Systems}},\ }\href
  {https://doi.org/10.1103/PhysRevX.4.041048} {\bibfield  {journal} {\bibinfo
  {journal} {Phys. Rev. X}\ }\textbf {\bibinfo {volume} {4}},\ \bibinfo {pages}
  {041048} (\bibinfo {year} {2014})}\BibitemShut {NoStop}%
\bibitem [{\citenamefont {Lazarides}\ \emph {et~al.}(2014)\citenamefont
  {Lazarides}, \citenamefont {Das},\ and\ \citenamefont
  {Moessner}}]{PhysRevE.90.012110}%
  \BibitemOpen
  \bibfield  {author} {\bibinfo {author} {\bibfnamefont {A.}~\bibnamefont
  {Lazarides}}, \bibinfo {author} {\bibfnamefont {A.}~\bibnamefont {Das}},\
  and\ \bibinfo {author} {\bibfnamefont {R.}~\bibnamefont {Moessner}},\
  }\bibfield  {title} {\bibinfo {title} {{Equilibrium states of generic quantum
  systems subject to periodic driving}},\ }\href
  {https://doi.org/10.1103/PhysRevE.90.012110} {\bibfield  {journal} {\bibinfo
  {journal} {Phys. Rev. E}\ }\textbf {\bibinfo {volume} {90}},\ \bibinfo
  {pages} {012110} (\bibinfo {year} {2014})}\BibitemShut {NoStop}%
\bibitem [{\citenamefont {{Abanin}}\ \emph {et~al.}(2015)\citenamefont
  {{Abanin}}, \citenamefont {{De Roeck}},\ and\ \citenamefont
  {{Huveneers}}}]{2015PhRvL.115y6803A}%
  \BibitemOpen
  \bibfield  {author} {\bibinfo {author} {\bibfnamefont {D.~A.}\ \bibnamefont
  {{Abanin}}}, \bibinfo {author} {\bibfnamefont {W.}~\bibnamefont {{De
  Roeck}}},\ and\ \bibinfo {author} {\bibfnamefont {F.}~\bibnamefont
  {{Huveneers}}},\ }\bibfield  {title} {\bibinfo {title} {{Exponentially Slow
  Heating in Periodically Driven Many-Body Systems}},\ }\href
  {https://doi.org/10.1103/PhysRevLett.115.256803} {\bibfield  {journal}
  {\bibinfo  {journal} {Phys. Rev. Lett.}\ }\textbf {\bibinfo {volume} {115}},\
  \bibinfo {eid} {256803} (\bibinfo {year} {2015})}\BibitemShut {NoStop}%
\bibitem [{\citenamefont {Else}\ \emph {et~al.}(2017)\citenamefont {Else},
  \citenamefont {Bauer},\ and\ \citenamefont {Nayak}}]{PhysRevX.7.011026}%
  \BibitemOpen
  \bibfield  {author} {\bibinfo {author} {\bibfnamefont {D.~V.}\ \bibnamefont
  {Else}}, \bibinfo {author} {\bibfnamefont {B.}~\bibnamefont {Bauer}},\ and\
  \bibinfo {author} {\bibfnamefont {C.}~\bibnamefont {Nayak}},\ }\bibfield
  {title} {\bibinfo {title} {{Prethermal Phases of Matter Protected by
  Time-Translation Symmetry}},\ }\href
  {https://doi.org/10.1103/PhysRevX.7.011026} {\bibfield  {journal} {\bibinfo
  {journal} {Phys. Rev. X}\ }\textbf {\bibinfo {volume} {7}},\ \bibinfo {pages}
  {011026} (\bibinfo {year} {2017})}\BibitemShut {NoStop}%
\bibitem [{\citenamefont {Bukov}\ \emph {et~al.}(2016)\citenamefont {Bukov},
  \citenamefont {Heyl}, \citenamefont {Huse},\ and\ \citenamefont
  {Polkovnikov}}]{PhysRevB.93.155132}%
  \BibitemOpen
  \bibfield  {author} {\bibinfo {author} {\bibfnamefont {M.}~\bibnamefont
  {Bukov}}, \bibinfo {author} {\bibfnamefont {M.}~\bibnamefont {Heyl}},
  \bibinfo {author} {\bibfnamefont {D.~A.}\ \bibnamefont {Huse}},\ and\
  \bibinfo {author} {\bibfnamefont {A.}~\bibnamefont {Polkovnikov}},\
  }\bibfield  {title} {\bibinfo {title} {{Heating and many-body resonances in a
  periodically driven two-band system}},\ }\href
  {https://doi.org/10.1103/PhysRevB.93.155132} {\bibfield  {journal} {\bibinfo
  {journal} {Phys. Rev. B}\ }\textbf {\bibinfo {volume} {93}},\ \bibinfo
  {pages} {155132} (\bibinfo {year} {2016})}\BibitemShut {NoStop}%
\bibitem [{\citenamefont {{Abanin}}\ \emph {et~al.}(2017)\citenamefont
  {{Abanin}}, \citenamefont {{De Roeck}}, \citenamefont {{Ho}},\ and\
  \citenamefont {{Huveneers}}}]{2017CMaPh.354..809A}%
  \BibitemOpen
  \bibfield  {author} {\bibinfo {author} {\bibfnamefont {D.}~\bibnamefont
  {{Abanin}}}, \bibinfo {author} {\bibfnamefont {W.}~\bibnamefont {{De
  Roeck}}}, \bibinfo {author} {\bibfnamefont {W.~W.}\ \bibnamefont {{Ho}}},\
  and\ \bibinfo {author} {\bibfnamefont {F.}~\bibnamefont {{Huveneers}}},\
  }\bibfield  {title} {\bibinfo {title} {{A Rigorous Theory of Many-Body
  Prethermalization for Periodically Driven and Closed Quantum Systems}},\
  }\href {https://doi.org/10.1007/s00220-017-2930-x} {\bibfield  {journal}
  {\bibinfo  {journal} {Communications in Mathematical Physics}\ }\textbf
  {\bibinfo {volume} {354}},\ \bibinfo {pages} {809} (\bibinfo {year}
  {2017})}\BibitemShut {NoStop}%
\bibitem [{\citenamefont {Luitz}\ \emph {et~al.}(2020)\citenamefont {Luitz},
  \citenamefont {Moessner}, \citenamefont {Sondhi},\ and\ \citenamefont
  {Khemani}}]{PhysRevX.10.021046}%
  \BibitemOpen
  \bibfield  {author} {\bibinfo {author} {\bibfnamefont {D.~J.}\ \bibnamefont
  {Luitz}}, \bibinfo {author} {\bibfnamefont {R.}~\bibnamefont {Moessner}},
  \bibinfo {author} {\bibfnamefont {S.~L.}\ \bibnamefont {Sondhi}},\ and\
  \bibinfo {author} {\bibfnamefont {V.}~\bibnamefont {Khemani}},\ }\bibfield
  {title} {\bibinfo {title} {{Prethermalization without Temperature}},\ }\href
  {https://doi.org/10.1103/PhysRevX.10.021046} {\bibfield  {journal} {\bibinfo
  {journal} {Phys. Rev. X}\ }\textbf {\bibinfo {volume} {10}},\ \bibinfo
  {pages} {021046} (\bibinfo {year} {2020})}\BibitemShut {NoStop}%
\bibitem [{\citenamefont {Lazarides}\ \emph {et~al.}(2015)\citenamefont
  {Lazarides}, \citenamefont {Das},\ and\ \citenamefont
  {Moessner}}]{PhysRevLett.115.030402}%
  \BibitemOpen
  \bibfield  {author} {\bibinfo {author} {\bibfnamefont {A.}~\bibnamefont
  {Lazarides}}, \bibinfo {author} {\bibfnamefont {A.}~\bibnamefont {Das}},\
  and\ \bibinfo {author} {\bibfnamefont {R.}~\bibnamefont {Moessner}},\
  }\bibfield  {title} {\bibinfo {title} {{Fate of Many-Body Localization Under
  Periodic Driving}},\ }\href {https://doi.org/10.1103/PhysRevLett.115.030402}
  {\bibfield  {journal} {\bibinfo  {journal} {Phys. Rev. Lett.}\ }\textbf
  {\bibinfo {volume} {115}},\ \bibinfo {pages} {030402} (\bibinfo {year}
  {2015})}\BibitemShut {NoStop}%
\bibitem [{\citenamefont {{Ponte}}\ \emph {et~al.}(2015)\citenamefont
  {{Ponte}}, \citenamefont {{Papi{\'c}}}, \citenamefont {{Huveneers}},\ and\
  \citenamefont {{Abanin}}}]{2015PhRvL.114n0401P}%
  \BibitemOpen
  \bibfield  {author} {\bibinfo {author} {\bibfnamefont {P.}~\bibnamefont
  {{Ponte}}}, \bibinfo {author} {\bibfnamefont {Z.}~\bibnamefont
  {{Papi{\'c}}}}, \bibinfo {author} {\bibfnamefont {F.}~\bibnamefont
  {{Huveneers}}},\ and\ \bibinfo {author} {\bibfnamefont {D.~A.}\ \bibnamefont
  {{Abanin}}},\ }\bibfield  {title} {\bibinfo {title} {{Many-Body Localization
  in Periodically Driven Systems}},\ }\href
  {https://doi.org/10.1103/PhysRevLett.114.140401} {\bibfield  {journal}
  {\bibinfo  {journal} {Phys. Rev. Lett.}\ }\textbf {\bibinfo {volume} {114}},\
  \bibinfo {eid} {140401} (\bibinfo {year} {2015})}\BibitemShut {NoStop}%
\bibitem [{\citenamefont {{Abanin}}\ \emph {et~al.}(2016)\citenamefont
  {{Abanin}}, \citenamefont {{De Roeck}},\ and\ \citenamefont
  {{Huveneers}}}]{2016AnPhy.372....1A}%
  \BibitemOpen
  \bibfield  {author} {\bibinfo {author} {\bibfnamefont {D.~A.}\ \bibnamefont
  {{Abanin}}}, \bibinfo {author} {\bibfnamefont {W.}~\bibnamefont {{De
  Roeck}}},\ and\ \bibinfo {author} {\bibfnamefont {F.}~\bibnamefont
  {{Huveneers}}},\ }\bibfield  {title} {\bibinfo {title} {{Theory of many-body
  localization in periodically driven systems}},\ }\href
  {https://doi.org/10.1016/j.aop.2016.03.010} {\bibfield  {journal} {\bibinfo
  {journal} {Annals of Physics}\ }\textbf {\bibinfo {volume} {372}},\ \bibinfo
  {pages} {1} (\bibinfo {year} {2016})}\BibitemShut {NoStop}%
\bibitem [{\citenamefont {Rubio-Abadal}\ \emph {et~al.}(2020)\citenamefont
  {Rubio-Abadal}, \citenamefont {Ippoliti}, \citenamefont {Hollerith},
  \citenamefont {Wei}, \citenamefont {Rui}, \citenamefont {Sondhi},
  \citenamefont {Khemani}, \citenamefont {Gross},\ and\ \citenamefont
  {Bloch}}]{PhysRevX.10.021044}%
  \BibitemOpen
  \bibfield  {author} {\bibinfo {author} {\bibfnamefont {A.}~\bibnamefont
  {Rubio-Abadal}}, \bibinfo {author} {\bibfnamefont {M.}~\bibnamefont
  {Ippoliti}}, \bibinfo {author} {\bibfnamefont {S.}~\bibnamefont {Hollerith}},
  \bibinfo {author} {\bibfnamefont {D.}~\bibnamefont {Wei}}, \bibinfo {author}
  {\bibfnamefont {J.}~\bibnamefont {Rui}}, \bibinfo {author} {\bibfnamefont
  {S.~L.}\ \bibnamefont {Sondhi}}, \bibinfo {author} {\bibfnamefont
  {V.}~\bibnamefont {Khemani}}, \bibinfo {author} {\bibfnamefont
  {C.}~\bibnamefont {Gross}},\ and\ \bibinfo {author} {\bibfnamefont
  {I.}~\bibnamefont {Bloch}},\ }\bibfield  {title} {\bibinfo {title} {{Floquet
  Prethermalization in a Bose-Hubbard System}},\ }\href
  {https://doi.org/10.1103/PhysRevX.10.021044} {\bibfield  {journal} {\bibinfo
  {journal} {Phys. Rev. X}\ }\textbf {\bibinfo {volume} {10}},\ \bibinfo
  {pages} {021044} (\bibinfo {year} {2020})}\BibitemShut {NoStop}%
\bibitem [{\citenamefont {Peng}\ \emph {et~al.}(2021)\citenamefont {Peng},
  \citenamefont {Yin}, \citenamefont {Huang}, \citenamefont {Ramanathan},\ and\
  \citenamefont {Cappellaro}}]{Peng:2021aa}%
  \BibitemOpen
  \bibfield  {author} {\bibinfo {author} {\bibfnamefont {P.}~\bibnamefont
  {Peng}}, \bibinfo {author} {\bibfnamefont {C.}~\bibnamefont {Yin}}, \bibinfo
  {author} {\bibfnamefont {X.}~\bibnamefont {Huang}}, \bibinfo {author}
  {\bibfnamefont {C.}~\bibnamefont {Ramanathan}},\ and\ \bibinfo {author}
  {\bibfnamefont {P.}~\bibnamefont {Cappellaro}},\ }\bibfield  {title}
  {\bibinfo {title} {Floquet prethermalization in dipolar spin chains},\
  }\href{https://doi.org/10.1038/s41567-020-01120-z} {\bibfield  {journal} {\bibinfo  {journal} {Nature Physics}\
  }\textbf {\bibinfo {volume} {17}},\ \bibinfo {pages} {444} (\bibinfo {year}
  {2021})}\BibitemShut {NoStop}%
\bibitem [{\citenamefont {{Basko}}\ \emph {et~al.}(2006)\citenamefont
  {{Basko}}, \citenamefont {{Aleiner}},\ and\ \citenamefont
  {{Altshuler}}}]{2006AnPhy.321.1126B}%
  \BibitemOpen
  \bibfield  {author} {\bibinfo {author} {\bibfnamefont {D.~M.}\ \bibnamefont
  {{Basko}}}, \bibinfo {author} {\bibfnamefont {I.~L.}\ \bibnamefont
  {{Aleiner}}},\ and\ \bibinfo {author} {\bibfnamefont {B.~L.}\ \bibnamefont
  {{Altshuler}}},\ }\bibfield  {title} {\bibinfo {title} {{Metal insulator
  transition in a weakly interacting many-electron system with localized
  single-particle states}},\ }\href {https://doi.org/10.1016/j.aop.2005.11.014}
  {\bibfield  {journal} {\bibinfo  {journal} {Annals of Physics}\ }\textbf
  {\bibinfo {volume} {321}},\ \bibinfo {pages} {1126} (\bibinfo {year}
  {2006})}\BibitemShut {NoStop}%
\bibitem [{\citenamefont {Smith}\ \emph {et~al.}(2017)\citenamefont {Smith},
  \citenamefont {Knolle}, \citenamefont {Kovrizhin},\ and\ \citenamefont
  {Moessner}}]{PhysRevLett.118.266601}%
  \BibitemOpen
  \bibfield  {author} {\bibinfo {author} {\bibfnamefont {A.}~\bibnamefont
  {Smith}}, \bibinfo {author} {\bibfnamefont {J.}~\bibnamefont {Knolle}},
  \bibinfo {author} {\bibfnamefont {D.~L.}\ \bibnamefont {Kovrizhin}},\ and\
  \bibinfo {author} {\bibfnamefont {R.}~\bibnamefont {Moessner}},\ }\bibfield
  {title} {\bibinfo {title} {{Disorder-Free Localization}},\ }\href
  {https://doi.org/10.1103/PhysRevLett.118.266601} {\bibfield  {journal}
  {\bibinfo  {journal} {Phys. Rev. Lett.}\ }\textbf {\bibinfo {volume} {118}},\
  \bibinfo {pages} {266601} (\bibinfo {year} {2017})}\BibitemShut {NoStop}%
\bibitem [{\citenamefont {Michielsen}\ and\ \citenamefont
  {De~Raedt}(1997)}]{Michielsen1997}%
  \BibitemOpen
  \bibfield  {author} {\bibinfo {author} {\bibfnamefont {K.}~\bibnamefont
  {Michielsen}}\ and\ \bibinfo {author} {\bibfnamefont {H.}~\bibnamefont
  {De~Raedt}},\ }\bibfield  {title} {\bibinfo {title} {{Quantum molecular
  dynamics study of the Su-Schrieffer-Heeger model}},\ }\href
  {https://doi.org/10.1007/s002570050393} {\bibfield  {journal} {\bibinfo
  {journal} {Z. Phys. B Condens. Mat.}\ }\textbf {\bibinfo {volume} {103}},\
  \bibinfo {pages} {391} (\bibinfo {year} {1997})}\BibitemShut {NoStop}%
\bibitem [{\citenamefont {Weber}\ \emph {et~al.}(2016)\citenamefont {Weber},
  \citenamefont {Assaad},\ and\ \citenamefont
  {Hohenadler}}]{PhysRevB.94.155150}%
  \BibitemOpen
  \bibfield  {author} {\bibinfo {author} {\bibfnamefont {M.}~\bibnamefont
  {Weber}}, \bibinfo {author} {\bibfnamefont {F.~F.}\ \bibnamefont {Assaad}},\
  and\ \bibinfo {author} {\bibfnamefont {M.}~\bibnamefont {Hohenadler}},\
  }\bibfield  {title} {\bibinfo {title} {{Thermodynamic and spectral properties
  of adiabatic Peierls chains}},\ }\href
  {https://doi.org/10.1103/PhysRevB.94.155150} {\bibfield  {journal} {\bibinfo
  {journal} {Phys. Rev. B}\ }\textbf {\bibinfo {volume} {94}},\ \bibinfo
  {pages} {155150} (\bibinfo {year} {2016})}\BibitemShut {NoStop}%
\bibitem [{\citenamefont {Rotvig}\ \emph {et~al.}(1995)\citenamefont {Rotvig},
  \citenamefont {Jauho},\ and\ \citenamefont {Smith}}]{PhysRevLett.74.1831}%
  \BibitemOpen
  \bibfield  {author} {\bibinfo {author} {\bibfnamefont {J.}~\bibnamefont
  {Rotvig}}, \bibinfo {author} {\bibfnamefont {A.-P.}\ \bibnamefont {Jauho}},\
  and\ \bibinfo {author} {\bibfnamefont {H.}~\bibnamefont {Smith}},\ }\bibfield
   {title} {\bibinfo {title} {{Bloch Oscillations, Zener Tunneling, and
  Wannier-Stark Ladders in the Time Domain}},\ }\href
  {https://doi.org/10.1103/PhysRevLett.74.1831} {\bibfield  {journal} {\bibinfo
   {journal} {Phys. Rev. Lett.}\ }\textbf {\bibinfo {volume} {74}},\ \bibinfo
  {pages} {1831} (\bibinfo {year} {1995})}\BibitemShut {NoStop}%
\bibitem [{\citenamefont {Bertoncini}\ and\ \citenamefont
  {Jauho}(1991)}]{PhysRevB.44.3655}%
  \BibitemOpen
  \bibfield  {author} {\bibinfo {author} {\bibfnamefont {R.}~\bibnamefont
  {Bertoncini}}\ and\ \bibinfo {author} {\bibfnamefont {A.~P.}\ \bibnamefont
  {Jauho}},\ }\bibfield  {title} {\bibinfo {title} {{Gauge-invariant
  formulation of the intracollisional field effect including collisional
  broadening}},\ }\href {https://doi.org/10.1103/PhysRevB.44.3655} {\bibfield
  {journal} {\bibinfo  {journal} {Phys. Rev. B}\ }\textbf {\bibinfo {volume}
  {44}},\ \bibinfo {pages} {3655} (\bibinfo {year} {1991})}\BibitemShut
  {NoStop}%
\bibitem [{\citenamefont {Lu}\ and\ \citenamefont {Raz}(2017)}]{mpemba}%
  \BibitemOpen
  \bibfield  {author} {\bibinfo {author} {\bibfnamefont {Z.}~\bibnamefont
  {Lu}}\ and\ \bibinfo {author} {\bibfnamefont {O.}~\bibnamefont {Raz}},\
  }\bibfield  {title} {\bibinfo {title} {Nonequilibrium thermodynamics of the
  {M}arkovian {M}pemba effect and its inverse},\ }\href
  {https://doi.org/10.1073/pnas.1701264114} {\bibfield  {journal} {\bibinfo
  {journal} {Proc. Nat. Acad. Sci. (USA)}\ }\textbf {\bibinfo {volume} {114}},\
  \bibinfo {pages} {5083} (\bibinfo {year} {2017})}\BibitemShut {NoStop}%
\bibitem [{Note1()}]{Note1}%
  \BibitemOpen
  \bibinfo {note} {See Supplemental Material at \protect \url {http://link.aps.org/supplemental/10.1103/PhysRevLett.130.266401} for details
  on the zero-temperature solution, additional results, a discussion of interaction effects in electronic models coupled to static variables, Refs.~\cite
  {McCoyWu+1973,RevModPhys.75.1333,PhysRevLett.122.247402,2020arXiv200308252R,PhysRev.138.B979,PhysRevLett.122.130604},
  and data files for the results presented in this Letter.}\BibitemShut {Stop}%
\bibitem [{\citenamefont {Leo}\ and\ \citenamefont
  {MacKinnon}(1989)}]{Leo_1989}%
  \BibitemOpen
  \bibfield  {author} {\bibinfo {author} {\bibfnamefont {J.}~\bibnamefont
  {Leo}}\ and\ \bibinfo {author} {\bibfnamefont {A.}~\bibnamefont
  {MacKinnon}},\ }\bibfield  {title} {\bibinfo {title} {{Stark-Wannier states
  and Stark ladders in semiconductor superlattices}},\ }\href
  {https://doi.org/10.1088/0953-8984/1/8/007} {\bibfield  {journal} {\bibinfo
  {journal} {Journal of Physics: Condensed Matter}\ }\textbf {\bibinfo {volume}
  {1}},\ \bibinfo {pages} {1449} (\bibinfo {year} {1989})}\BibitemShut
  {NoStop}%
\bibitem [{\citenamefont {Schneider}\ \emph {et~al.}(1990)\citenamefont
  {Schneider}, \citenamefont {Grahn}, \citenamefont {Klitzing},\ and\
  \citenamefont {Ploog}}]{PhysRevLett.65.2720}%
  \BibitemOpen
  \bibfield  {author} {\bibinfo {author} {\bibfnamefont {H.}~\bibnamefont
  {Schneider}}, \bibinfo {author} {\bibfnamefont {H.~T.}\ \bibnamefont
  {Grahn}}, \bibinfo {author} {\bibfnamefont {K.~v.}\ \bibnamefont
  {Klitzing}},\ and\ \bibinfo {author} {\bibfnamefont {K.}~\bibnamefont
  {Ploog}},\ }\bibfield  {title} {\bibinfo {title} {{Resonance-induced
  delocalization of electrons in GaAs-AlAs superlattices}},\ }\href
  {https://doi.org/10.1103/PhysRevLett.65.2720} {\bibfield  {journal} {\bibinfo
   {journal} {Phys. Rev. Lett.}\ }\textbf {\bibinfo {volume} {65}},\ \bibinfo
  {pages} {2720} (\bibinfo {year} {1990})}\BibitemShut {NoStop}%
\bibitem [{\citenamefont {Prigodin}(1980)}]{Prigodin1980}%
  \BibitemOpen
  \bibfield  {author} {\bibinfo {author} {\bibfnamefont {V.~N.}\ \bibnamefont
  {Prigodin}},\ }\bibfield  {title} {\bibinfo {title} {One-dimensional
  disordered system in an electric field},\ }\href@noop {} {\bibfield
  {journal} {\bibinfo  {journal} {JETP}\ }\textbf {\bibinfo {volume} {52}},\
  \bibinfo {pages} {1185} (\bibinfo {year} {1980})}\BibitemShut {NoStop}%
\bibitem [{\citenamefont {Cota}\ \emph {et~al.}(1987)\citenamefont {Cota},
  \citenamefont {Jos\'e},\ and\ \citenamefont
  {Monsiv\'ais}}]{PhysRevB.35.8929}%
  \BibitemOpen
  \bibfield  {author} {\bibinfo {author} {\bibfnamefont {E.}~\bibnamefont
  {Cota}}, \bibinfo {author} {\bibfnamefont {J.~V.}\ \bibnamefont {Jos\'e}},\
  and\ \bibinfo {author} {\bibfnamefont {G.}~\bibnamefont {Monsiv\'ais}},\
  }\bibfield  {title} {\bibinfo {title} {{Stark-ladder resonances in ordered
  and disordered electrified chains}},\ }\href
  {https://doi.org/10.1103/PhysRevB.35.8929} {\bibfield  {journal} {\bibinfo
  {journal} {Phys. Rev. B}\ }\textbf {\bibinfo {volume} {35}},\ \bibinfo
  {pages} {8929} (\bibinfo {year} {1987})}\BibitemShut {NoStop}%
\bibitem [{Note2()}]{Note2}%
  \BibitemOpen
  \bibinfo {note} {In equilibrium, the adiabatic-phonon description is valid
  for temperatures much larger than the phonon frequency, as confirmed by exact
  quantum Monte Carlo simulations \cite {PhysRevB.98.235117}.}\BibitemShut
  {Stop}%
\bibitem [{\citenamefont {McCoy}\ and\ \citenamefont
  {Wu}(1973)}]{McCoyWu+1973}%
  \BibitemOpen
  \bibfield  {author} {\bibinfo {author} {\bibfnamefont {B.~M.}\ \bibnamefont
  {McCoy}}\ and\ \bibinfo {author} {\bibfnamefont {T.~T.}\ \bibnamefont {Wu}},\
  }\href {https://doi.org/doi:10.4159/harvard.9780674180758} {\emph {\bibinfo
  {title} {The Two-Dimensional Ising Model}}}\ (\bibinfo  {publisher} {Harvard
  University Press},\ \bibinfo {address} {Cambridge, MA and London, England},\
  \bibinfo {year} {1973})\BibitemShut {NoStop}%
\bibitem [{\citenamefont {Freericks}\ and\ \citenamefont
  {Zlati\ifmmode~\acute{c}\else \'{c}\fi{}}(2003)}]{RevModPhys.75.1333}%
  \BibitemOpen
  \bibfield  {author} {\bibinfo {author} {\bibfnamefont {J.~K.}\ \bibnamefont
  {Freericks}}\ and\ \bibinfo {author} {\bibfnamefont {V.}~\bibnamefont
  {Zlati\ifmmode~\acute{c}\else \'{c}\fi{}}},\ }\bibfield  {title} {\bibinfo
  {title} {{Exact dynamical mean-field theory of the Falicov-Kimball model}},\
  }\href {https://doi.org/10.1103/RevModPhys.75.1333} {\bibfield  {journal}
  {\bibinfo  {journal} {Rev. Mod. Phys.}\ }\textbf {\bibinfo {volume} {75}},\
  \bibinfo {pages} {1333} (\bibinfo {year} {2003})}\BibitemShut {NoStop}%
\bibitem [{\citenamefont {Matveev}\ \emph {et~al.}(2019)\citenamefont
  {Matveev}, \citenamefont {Shvaika}, \citenamefont {Devereaux},\ and\
  \citenamefont {Freericks}}]{PhysRevLett.122.247402}%
  \BibitemOpen
  \bibfield  {author} {\bibinfo {author} {\bibfnamefont {O.~P.}\ \bibnamefont
  {Matveev}}, \bibinfo {author} {\bibfnamefont {A.~M.}\ \bibnamefont
  {Shvaika}}, \bibinfo {author} {\bibfnamefont {T.~P.}\ \bibnamefont
  {Devereaux}},\ and\ \bibinfo {author} {\bibfnamefont {J.~K.}\ \bibnamefont
  {Freericks}},\ }\bibfield  {title} {\bibinfo {title} {Stroboscopic tests for
  thermalization of electrons in pump-probe experiments},\ }\href
  {https://doi.org/10.1103/PhysRevLett.122.247402} {\bibfield  {journal}
  {\bibinfo  {journal} {Phys. Rev. Lett.}\ }\textbf {\bibinfo {volume} {122}},\
  \bibinfo {pages} {247402} (\bibinfo {year} {2019})}\BibitemShut {NoStop}%
\bibitem [{\citenamefont {{Rudner}}\ and\ \citenamefont
  {{Lindner}}(2020{\natexlab{b}})}]{2020arXiv200308252R}%
  \BibitemOpen
  \bibfield  {author} {\bibinfo {author} {\bibfnamefont {M.~S.}\ \bibnamefont
  {{Rudner}}}\ and\ \bibinfo {author} {\bibfnamefont {N.~H.}\ \bibnamefont
  {{Lindner}}},\ }\bibfield  {title} {\bibinfo {title} {{The Floquet Engineer's
  Handbook}},\ }\href{https://doi.org/10.48550/arXiv.2003.08252} {\bibfield  {journal} {\bibinfo  {journal}
  {arXiv:2003.08252}} (\bibinfo {year} {2020}{\natexlab{b}})}\BibitemShut
  {NoStop}%
\bibitem [{\citenamefont {Shirley}(1965)}]{PhysRev.138.B979}%
  \BibitemOpen
  \bibfield  {author} {\bibinfo {author} {\bibfnamefont {J.~H.}\ \bibnamefont
  {Shirley}},\ }\bibfield  {title} {\bibinfo {title} {{Solution of the
  Schr\"odinger Equation with a Hamiltonian Periodic in Time}},\ }\href
  {https://doi.org/10.1103/PhysRev.138.B979} {\bibfield  {journal} {\bibinfo
  {journal} {Phys. Rev.}\ }\textbf {\bibinfo {volume} {138}},\ \bibinfo {pages}
  {B979} (\bibinfo {year} {1965})}\BibitemShut {NoStop}%
\bibitem [{\citenamefont {Uhrig}\ \emph {et~al.}(2019)\citenamefont {Uhrig},
  \citenamefont {Kalthoff},\ and\ \citenamefont
  {Freericks}}]{PhysRevLett.122.130604}%
  \BibitemOpen
  \bibfield  {author} {\bibinfo {author} {\bibfnamefont {G.~S.}\ \bibnamefont
  {Uhrig}}, \bibinfo {author} {\bibfnamefont {M.~H.}\ \bibnamefont
  {Kalthoff}},\ and\ \bibinfo {author} {\bibfnamefont {J.~K.}\ \bibnamefont
  {Freericks}},\ }\bibfield  {title} {\bibinfo {title} {{Positivity of the
  Spectral Densities of Retarded Floquet Green Functions}},\ }\href
  {https://doi.org/10.1103/PhysRevLett.122.130604} {\bibfield  {journal}
  {\bibinfo  {journal} {Phys. Rev. Lett.}\ }\textbf {\bibinfo {volume} {122}},\
  \bibinfo {pages} {130604} (\bibinfo {year} {2019})}\BibitemShut {NoStop}%
\bibitem [{\citenamefont {Weber}\ \emph {et~al.}(2018)\citenamefont {Weber},
  \citenamefont {Assaad},\ and\ \citenamefont
  {Hohenadler}}]{PhysRevB.98.235117}%
  \BibitemOpen
  \bibfield  {author} {\bibinfo {author} {\bibfnamefont {M.}~\bibnamefont
  {Weber}}, \bibinfo {author} {\bibfnamefont {F.~F.}\ \bibnamefont {Assaad}},\
  and\ \bibinfo {author} {\bibfnamefont {M.}~\bibnamefont {Hohenadler}},\
  }\bibfield  {title} {\bibinfo {title} {{Thermal and quantum lattice
  fluctuations in Peierls chains}},\ }\href
  {https://doi.org/10.1103/PhysRevB.98.235117} {\bibfield  {journal} {\bibinfo
  {journal} {Phys. Rev. B}\ }\textbf {\bibinfo {volume} {98}},\ \bibinfo
  {pages} {235117} (\bibinfo {year} {2018})}\BibitemShut {NoStop}%
\end{thebibliography}
\end{document}